\documentclass{aa}  

\usepackage{txfonts}

\usepackage{graphicx,rotating,amssymb,amsmath}
\usepackage{booktabs}
\usepackage{subfig}
\usepackage{float,capt-of}
\usepackage{natbib}
\usepackage{multirow}
\usepackage[english]{babel}
\usepackage{morefloats}
\usepackage{color}
\usepackage{xcolor}

\usepackage{enumitem}
\usepackage{hyperref}
\usepackage{multirow}
\usepackage{multicol}
\usepackage{ulem}

\newcommand{\sfr}{\mathrm{SFR}}
\newcommand{\msun}{M_\odot}

\newcommand{\ha}{H$\alpha$ }

\begin{document}
   \title{Higher resolution optical spectra of $M_*<10^{10}~M_{\odot}$ galaxies reveal outflow signatures unresolved by the SDSS}

   \author{B. Hagedorn \inst{1}
          \and
          C. Cicone \inst{1} 
          \and
          C. Vignali \inst{2,3}
          \and
          P. Severgnini \inst{4}
          \and
          M. Pedani \inst{5}
          \and
          M. Sarzi \inst{6}
          \and
          A. Saintonge \inst{7,8}
          \and
          M. Romano \inst{7,9}
          }
      \institute{\inst{1}Institute of Theoretical Astrophysics, University of Oslo, P.O. Box 1029, Blindern, 0315 Oslo, Norway\\ 
      \inst{2}Dipartimento di Fisica e Astronomia Augusto Righi, Università degli Studi di Bologna, via Gobetti 93/2, 40129 Bologna, Italy\\ 
      \inst{3}INAF – Osservatorio di Astrofisica e Scienza dello Spazio di Bologna, Via Gobetti 101, 40129 Bologna, Italy \\
      \inst{4}INAF – Osservatorio Astronomico di Brera, Via Brera 28, 20121 Milano, Italy \\ 
      \inst{5}INAF – Fundación Galileo Galilei, Rambla José Ana Fernandez Pérez 7, 38712 Breña Baja, TF, Spain \\ 
      \inst{6}Armagh Observatory and Planetarium, College Hill, Armagh BT61 9DG, UK \\ 
      \inst{7}Max-Planck-Institut f\"ur Radioastronomie, Auf dem H\"ugel 69, 53121 Bonn, Germany\\ 
      \inst{8}Department of Physics and Astronomy, University College London, London, WC1E 6BT, UK \\
      \inst{9}INAF – Osservatorio Astronomico di Padova, Vicolo dell’Osservatorio 5, 35122, Padova, Italy\\
      \email{bendix.hagedorn@astro.uio.no}
		      }
		
   \date{Received: 19 December 2025 / Accepted: 1 April 2026}


\abstract{
Galactic outflows are predicted to be ubiquitous in low-mass galaxies, but observational evidence is lacking. 
Both a low signal-to-noise and a low spectral resolution can severely hamper the detection of galactic outflows, especially in small galaxies that have intrinsically narrow spectral lines.
In an effort to overcome these issues, we obtained new, medium-high resolution (FWHM$_\mathrm{inst}\sim50-110$\,km/s) optical spectra of 52 local star forming galaxies ($0.01 < z < 0.03$) with stellar masses $10^{8.5}<M_*/[M_{\odot}]<10^{10}$, using the TNG/DOLORES and NTT/EFOSC2 instruments. 
Our parent sample consists of SDSS galaxies with available heterodyne single-dish molecular (i.e., CO) line data. The targets of this study are selected among those that, based on the comparison between CO line widths, SDSS spectral resolution, and corresponding SDSS-based H$\alpha$ line widths, have a high chance of being unresolved by SDSS spectroscopy.
Our new, medium-high resolution spectra reveal overall narrower \ha and [OIII]$\lambda5007$ lines, with signs of asymmetries and broad wings that are absent in the SDSS spectra of the same galaxies. 
This confirms our hypothesis that SDSS spectroscopy does not resolve the narrow emission lines of low-M$_*$ galaxies, which hinders the detection of outflows.
We identify outflow signatures in $\sim30\%$ of our targets based on the \ha line spectra.
Assuming a typical bi-conical outflow geometry, this detection rate is consistent with theoretical predictions of ubiquitous outflows in the low-mass regime.
The outflow incidence is enhanced ($\sim60\%$) for galaxies with above average star formation rates for the sample (SFR $>10^{-0.74}\,\mathrm{\msun/yr}$).
We estimate ionized gas mass outflow rates ranging from $\sim0.1 - 50\times10^{-3}\,\mathrm{\msun/yr}$ (mean $\sim20\times10^{-3}\,\mathrm{\msun/yr}$) and corresponding mass loading factors between 0.03 and 0.14 (mean $\sim0.07$) for the sample.
}


\titlerunning{Galactic outflows in low-mass galaxies}
\authorrunning{B. Hagedorn et al.}
\maketitle
%
\section{Introduction}
\label{sec:intro}
In the $\Lambda$CDM cosmological model, galaxies grow within a network of dark matter filaments, through a finely tuned exchange of material between their interstellar medium (ISM), the surrounding circum-galactic medium (CGM), and the external environment. 
Galactic outflows are thought to be key actors of this “baryon cycle”, as they can transport gas and dust away from galaxy nuclei and disks, possibly carrying some beyond the virial radius hence depleting the ISM and regulating star formation. 
This is likely instrumental in `quenching', i.e. the transition of galaxies from star forming to quiescent \citep{veilleuxGalacticWinds2005, veilleuxCoolOutflowsGalaxies2020}. The exact physical mechanisms behind quenching remains one of the central unsolved questions in the theory of galaxy evolution. 

Outflows are generally multi-phase \citep[e.g.,][]{ciconeLargelyUnconstrainedMultiphase2018}, requiring multi-wavelengths observations to fully characterize.
Observations of a single phase, such as those of the warm ionized phase we present here, cannot be used to fully constrain mass outflow rates for instance, particularly since the colder, denser phases frequently dominate the outflow mass \citep[e.g.,][]{carnianiIonisedOutflows242015, fluetschPropertiesMultiphaseOutflows2021, Hagedorn2026MassiveMultiphase}.
Nonetheless, they are useful to determine the incidence of outflows and inform follow-up observations at complementary wavelengths.

Theoretical models predict that quenching due to outflows driven by star formation is most effective in low-mass galaxies, as gas is more easily ejected from their shallow gravitational potentials \citep{dekelOriginDwarfGalaxies1986, lowStarburstdrivenMassLoss1999, ferraraRoleStellarFeedback2000, veilleuxGalacticWinds2005, daveGalaxyEvolutionCosmological2011}.
This has been suggested as an explanation for the decreasing stellar mass to halo mass relation in the low-mass regime \citep{behrooziCOMPREHENSIVEANALYSISUNCERTAINTIES2010}.
In addition, outflows are predicted to effectively remove metals from the ISM of low-mass galaxies \citep{christensenTracingOutflowingMetals2018}, offering one possible explanation for the low observed fractions of metals retained in these objects \citep{mcquinnLEOHOWMANY2015} and the resulting mass-metallicity relation \citep{brooksOriginEvolutionMassMetallicity2007}.
Outflows have also been invoked as an explanation for the `cored' dark matter halos of dwarf galaxies \citep[][]{navarroCoresDwarfGalaxy1996, governatoBulgelessDwarfGalaxies2010, governatoCuspyNoMore2012, teyssierCuspcoreTransformationsDwarf2013}.

All this suggests that outflows should be near ubiquitous in the low-mass regime.
Such outflows have indeed been observed in ionized gas tracers \citep[e.g.,][]{manzano-kingAGNDrivenOutflowsDwarf2019, mcquinnGalacticWindsLowmass2019, xuEMPRESSVIOutflows2022, marascoShakenNotExpelled2023}.
However, some studies covering wide mass ranges report outflow incidence declining with stellar mass \citep[e.g.,][]{ciconeOutflowsComplexStellar2016, concasLightBreezeLocal2017, averyIncidenceScalingRelations2021}.
This may be due to the fact that detecting outflow signatures requires high quality data, and spectroscopic galaxy surveys at optical wavelengths are generally not optimized for outflow identification.
Commonly used techniques for identifying outflows rely on analyzing the kinematics of the gas through observations of spectral lines in emission or absorption \citep[e.g.,][]{rupkeINTEGRALFIELDSPECTROSCOPY2011,ciconeMassiveMolecularOutflows2014}.
Outflows produce signatures in the profiles of spectral lines in the form of faint, possibly asymmetric, high-velocity features, distinct from those produced by ISM components in gravitationally dominated motion.
Detecting faint features naturally requires high signal-to-noise spectra, while deviations from line profiles produced by rotating disks can only be seen in spectra with sufficient spectral resolution to resolve such details of the line profile.
The former issue can be overcome to some extent by spectral stacking \citep[e.g.][]{ciconeOutflowsComplexStellar2016, concasLightBreezeLocal2017}, but the latter requires the use of appropriate instruments, especially in the case of low-mass galaxies which have overall narrower line profiles, which may not be resolved in optical spectra from galaxy surveys such as the SDSS.
It is therefore somewhat unclear whether the low incidence of outflows in the low-mass regime stands in contrast to theoretical predictions or whether it is due to the lack of quality data in this regime.

In this work, we present medium-high resolution resolution optical spectroscopy of 52 local ($0.01<z<0.03$) galaxies with stellar masses ($M_*$) between $10^{8.5}\,\mathrm{M_\odot}$ and $10^{10}\,\mathrm{M_\odot}$, and with emission lines unresolved by SDSS spectroscopy (hereafter we refer to the new data simply as high-resolution).
This sample is selected from galaxies observed in carbon monoxide (CO(1-0) or CO(2-1)) emission with single-dish heterodyne instruments on the APEX and IRAM telescopes as part of the ALLSMOG and xCOLDGASS surveys.
The very high spectral resolution of the molecular line observations allows us to identify candidates that are potentially unresolved in optical SDSS spectra (see Section~\ref{sec:sample} for details on the selection process).

The paper is structured as follows.
In Section~\ref{sec:sample}, we outline the sample selection.
In Section~\ref{sec:data}, we describe the observations and data reduction process.
Section~\ref{sec:methods} describes the methods used in fitting the emission line profile and identifying the outflows.
In Section~\ref{sec:results}, we show the effect of the improved spectral resolution and and derive the outflow incidence for the sample.
In Section~\ref{sec:discussion}, we discuss the implications of our results with respect to previous observations and theoretical predictions.
In Section~\ref{sec:conclusion}, we offer a brief summary and conclude with the main findings of this work.
Throughout this paper we assume a flat $\mathrm{\Lambda CDM}$ cosmology with parameters from the \citet{planckcollaborationPlanck2018Results2020} ($H_0=67.7\,\mathrm{km\,s^{-1}\,Mpc^{-1}}$, $\Omega_m=0.31$, $\Omega_\Lambda=0.69$) for cosmological distance calculations.

\section{Target selection}
\label{sec:sample}
Our targets are drawn from a parent sample sample of galaxies with available molecular line coverage, specifically from the single-dish ALLSMOG \citep{bothwellALLSMOGAPEXLowredshift2014, ciconeFinalDataRelease2017, hagedornMolecularGasScaling2024} and xCOLDGASS \citep{saintongeXCOLDGASSComplete2017} surveys, and photometry and spectroscopy from SDSS DR8 \citep{AiharaSDSSdr8}.
Below, we explain our sample selection, which relies on a comparison between CO and H$\alpha$ line widths.

\begin{figure}
    \centering
    \includegraphics[clip=true, trim=0.4cm 0.3cm 0.4cm 0.4cm,width=0.49\textwidth]{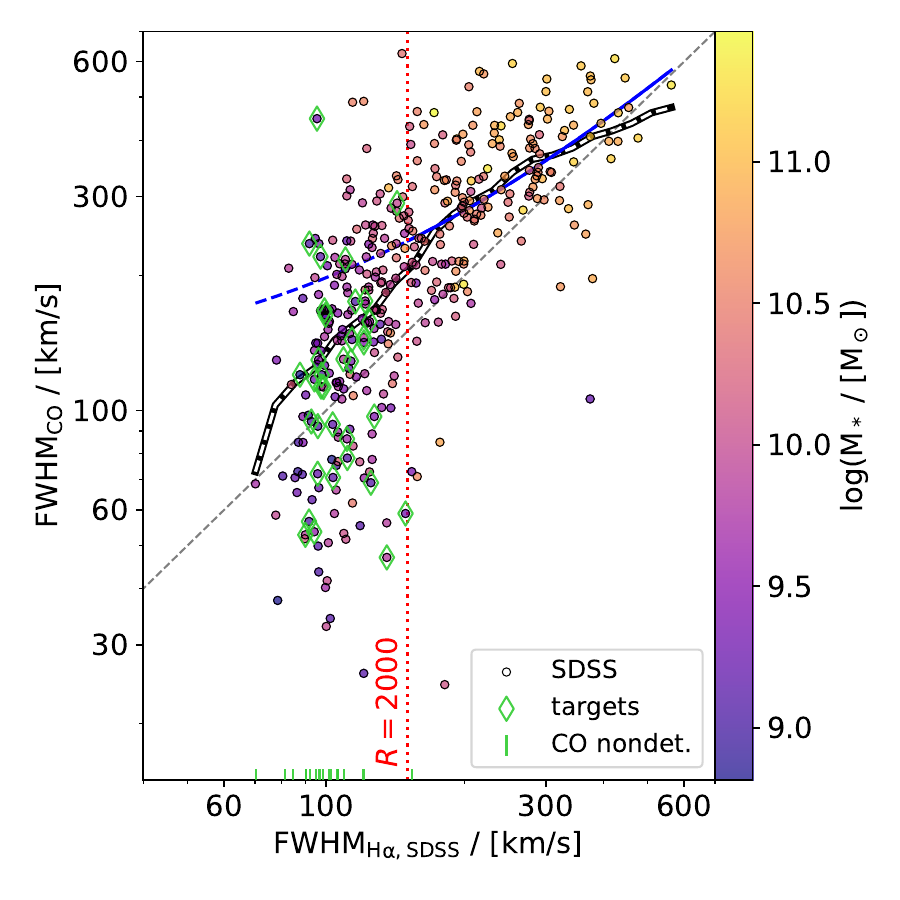}
    \caption{FWHM of the H$\alpha$ line from SDSS DR8 vs FWHM of CO from ALLSMOG and xCOLDGASS. The red dotted line shows the velocity resolution equivalent of $R=2000$ (FWHM $\sim150$\,km/s), which is typical for SDSS DR8 at this redshift.
    The gray dashed line shows the one-to-one relation.
    Galaxies targeted with high-resolution spectroscopy are marked with green diamonds.
    Targets without detections in CO are marked with vertical line markers on the x-axis at the position corresponding to their SDSS-inferred \ha line widths.
    The black-and-white dashed line shows the running median in logarithmic bins of 0.2\,dex width.
    The solid blue line shows a linear fit to the data with resolved \ha lines (i.e., to the right of the $R=2000$ line), which is extrapolated into to unresolved regime as a dashed line. The color scaling corresponds to the stellar mass of the galaxies.}
    \label{fig:allgass_ha_co_fwhm}
\end{figure}

Figure~\ref{fig:allgass_ha_co_fwhm} shows the H$\alpha$ emission line widths obtained from the SDSS spectra (using a single-Gaussian fit accounting for instrumental resolution, see Section~\ref{sec:methods:line_fitting}) as a function of the CO line widths adopted from ALLSMOG and xCOLDGASS for all CO-detections in the respective samples.
For galaxies with line widths larger than the SDSS spectral resolution (i.e., to the right of the vertical red dashed line in Fig.~\ref{fig:allgass_ha_co_fwhm}) there is a correlation between \ha and CO line widths. 
We also note that, in this regime, CO appears systematically broader than H$\alpha$.
This is likely due to the 3'' SDSS fiber only probing the center of these galaxies, while the CO observations cover a much larger part of the disk (see further discussion in Section~\ref{sec:res:ha_co_comp}).
To the left of the vertical red dashed line (i.e., below the SDSS spectral resolution) the \ha line widths inferred from SDSS spectroscopy scatter around a value of $\sim100$\,km/s regardless of the associated CO line widths, and the correlation between \ha and CO line widths breaks down. 
This is further illustrated by the deviation between the binned median line width (black line) and the extrapolation to lower FWHM values of the linear fit performed on the spectrally resolved (higher-FWHM) data (dashed blue line in Fig.~\ref{fig:allgass_ha_co_fwhm}).
This is clearly a resolution effect, and it strongly indicates that some of the $M_*\lesssim10^{10}\,M_{\odot}$ galaxies are unresolved by SDSS spectroscopy and so their H$\alpha$ line width measurements are dominated by the instrumental dispersion and associated uncertainties.
We confirm this in Appendix~\ref{app:linewidths}, where we compare the line widths obtained from SDSS spectra directly to those obtained from higher resolution spectra.

The goal of this study is to test whether the low SDSS resolution could hide ionized outflow signatures in low-M$_*$ galaxies.
The targets of our TNG/DOLORES and NTT/EFOSC2 follow up that are detected in CO are marked in Fig.~\ref{fig:allgass_ha_co_fwhm} using green diamonds. 
To avoid biases towards especially gas-rich galaxies, our sample also includes galaxies without CO detections that span the same range of SDSS-based inferred \ha line widths.
Targets were selected to have $M_*<10^{10}\,M_{\odot}$, $z<0.05$, and SDSS based \ha line widths below the SDSS resolution.
For observations with TNG/DOLORES, we further restrict targets to have $\mathrm{mag_R}<17.5$ and $\mathrm{mag_V}<17$.
For observations with NTT/EFOSC2 magnitudes were restricted to $\mathrm{mag_R}<18.5$ and $\mathrm{mag_V}<18.5$.
Magnitudes were calculated based on r- and g-band magnitudes measured in the SDSS fiber.
We adopt stellar masses from the MPA-JHU value added catalog\footnote{\url{https://www.sdss4.org/dr17/spectro/galaxy_mpajhu/}} for SDSS DR8 and redshifts from ALLSMOG and xCOLDGASS.
Based on these criteria, targets were selected to fill the allocated observing nights efficiently and minimize the air mass during observations.
Due to weather induced shifts in the schedule for the NTT/EFOSC2 observations, we observed some additional targets not in the originally selected sample, some of which had R- and V-band magnitudes between 18.5 and 19 (2MASSJ00114318+1428010, LEDA1397680, and UGC11992).
The full target list of 52 galaxies is reported in Table~\ref{tab:target_info}.

Our sample ranges in stellar mass from $\sim10^{8.5}\,\mathrm{M_\odot}$ and $\sim10^{10}\,\mathrm{M_\odot}$ and in SFR from $\sim10^{-2.5}\,\mathrm{M_\odot/yr}$ and $\sim1\,\mathrm{M_\odot/yr}$.
The majority of galaxies in the sample are star-forming based on the criterion defined in \citet{hagedornMolecularGasScaling2024} using the offset from the main sequence of star forming galaxies as defined by \citet{saintongeColdInterstellarMedium2022}.
Four targets (SDSSJ011437.36+011053.4, SDSSJ120108.03+152353.6, UGC11992, and LEDA213877) lie far enough below the main sequence to be defined as quiescent using this definition.
Using emission line diagnostics \citep{baldwinClassificationParametersEmissionline1981,kewleyTheoreticalModelingStarburst2001, kauffmannHostGalaxiesActive2003}, one galaxy in the sample (SDSSJ120108.03+152353.6) shows AGN-like ionization, and one (NGC2607) is classified as LINER.
The remaining galaxies show star-formation like ionization, except for LEDA213877, which could not be classified due to its lack of detected emission lines in both SDSS and TNG spectra.

\section{Data}
\label{sec:data}
We obtained high resolution spectroscopy of 52 local galaxies.
Targets are listed in Table~\ref{tab:target_info} and the emission line spectra used in this work are shown in Appendices \ref{app:ntt_specs} and \ref{app:tng_specs}.

\begin{table*}
\caption{\raggedright Target properties.}
    \centering
    \begin{tabular}{lccccccc}
    \hline
    \hline
Target & RA & Dec & z & log($M_*$) & log(SFR) & FWHM$_\mathrm{CO}$ & CO source\\
 & [deg] & [deg] & & $[\mathrm{M_\odot}]$ & $[\mathrm{M_\odot/yr}]$ & [km/s] & \\
\hline
2MASSJ00114318+1428010 & 2.930 & 14.467 & 0.0174 & $8.7\pm0.12$ & $-1\pm0.1$ & - & (1) \\
AGC244381 & 217.198 & 13.821 & 0.0176 & $9.2\pm0.08$ & $-2\pm0.3$ & $160$ & (2) \\
IC159 & 26.604 & -8.637 & 0.0131 & $9.7\pm0.07$ & $-1.7\pm0.1$ & $200\pm30$ & (1) \\
IC5292 & 348.446 & 13.688 & 0.0154 & $9.8\pm0.09$ & $-0.6\pm0.12$ & $130$ & (2) \\
LEDA1278497 & 166.061 & 5.127 & 0.0176 & $8.7\pm0.08$ & $-1.9\pm0.11$ & - & (1) \\
LEDA1397680 & 246.353 & 11.714 & 0.0163 & $9.1\pm0.13$ & $-1\pm0.09$ & - & (1) \\
LEDA1417541 & 246.062 & 12.856 & 0.0164 & $9.1\pm0.07$ & $-2\pm0.4$ & $200\pm120$ & (1) \\
LEDA1439867 & 167.623 & 13.766 & 0.0144 & $8.6\pm0.08$ & $-1.2\pm0.11$ & - & (1) \\
LEDA1446233 & 22.455 & 13.991 & 0.0233 & $9.1\pm0.09$ & $-2\pm0.2$ & $100\pm40$ & (1) \\
LEDA1452135 & 30.183 & 14.211 & 0.0160 & $8.9\pm0.08$ & $-2\pm0.3$ & - & (1) \\
LEDA1464874 & 30.267 & 14.701 & 0.0152 & $9\pm0.09$ & $-2\pm0.16$ & - & (1) \\
LEDA3091160 & 210.501 & 10.831 & 0.0131 & $9.1\pm0.09$ & $-0.9\pm0.09$ & $90$ & (2) \\
LEDA3127469 & 8.435 & 14.408 & 0.0179 & $8.6\pm0.07$ & $-2.1\pm0.11$ & - & (1) \\
LEDA3530 & 14.767 & 1.001 & 0.0178 & $8.8\pm0.06$ & $-0.8\pm0.09$ & - & (1) \\
LEDA49155 & 207.705 & 17.282 & 0.0175 & $9.2\pm0.08$ & $-1.5\pm0.1$ & $220$ & (2) \\
MCG+00-34-035 & 202.195 & -2.041 & 0.0124 & $8.8\pm0.13$ & $-1.6\pm0.04$ & $100\pm40$ & (1) \\
NGC7401 & 343.244 & 1.143 & 0.0164 & $9.5\pm0.08$ & $-2\pm0.2$ & $110$ & (2) \\
NGC99 & 5.997 & 15.770 & 0.0177 & $9.9\pm0.07$ & $-1.4\pm0.12$ & $110$ & (2) \\
SDSSJ002048.58+141327.8 & 5.202 & 14.224 & 0.0178 & $9.3\pm0.09$ & $-0.3\pm0.09$ & $60\pm16$ & (1) \\
SDSSJ011437.36+011053.4 & 18.656 & 1.181 & 0.0154 & $9.8\pm0.11$ & $-2\pm0.4$ & $50$ & (2) \\
SDSSJ120108.03+152353.6 & 180.283 & 15.398 & 0.0172 & $9.7\pm0.1$ & $-2\pm0.8$ & - & (2) \\
SDSSJ135906.39+175604.9 & 209.777 & 17.935 & 0.0176 & $9.1\pm0.09$ & $-1\pm0.1$ & $70$ & (2) \\
SDSSJ140317.75+100340.3 & 210.824 & 10.061 & 0.0134 & $8.5\pm0.1$ & $-1.3\pm0.09$ & - & (1) \\
UGC10005 & 236.310 & 0.772 & 0.0128 & $9.1\pm0.13$ & $-3\pm0.3$ & $100\pm20$ & (1) \\
UGC11992 & 335.197 & 14.235 & 0.0120 & $9.2\pm0.09$ & $-2\pm0.4$ & $70$ & (2) \\
UGC272 & 6.957 & -1.200 & 0.0130 & $9.4\pm0.07$ & $-2\pm0.2$ & - & (1) \\
UGC6838 & 177.984 & -2.642 & 0.0129 & $9.3\pm0.08$ & $-1.6\pm0.14$ & $400\pm70$ & (1) \\
UGC9359 & 218.316 & -1.140 & 0.0138 & $9\pm0.09$ & $-1.9\pm0.12$ & - & (1) \\
Z38-48 & 163.803 & 5.863 & 0.0117 & $9.2\pm0.09$ & $-1.6\pm0.12$ & $90$ & (2) \\
Z383-30 & 7.410 & 0.410 & 0.0138 & $9.2\pm0.07$ & $-2.1\pm0.14$ & $70$ & (2) \\
Z386-19 & 24.405 & 0.040 & 0.0164 & $9.3\pm0.08$ & $-1.5\pm0.15$ & $120$ & (2) \\
Z429-16 & 339.841 & 13.882 & 0.0173 & $9.1\pm0.16$ & $-0.7\pm0.08$ & $80$ & (2) \\
Z429-21 & 340.294 & 13.340 & 0.0173 & $9.5\pm0.07$ & $-1.3\pm0.13$ & $170$ & (2) \\
Z436-34 & 20.378 & 14.505 & 0.0140 & $9.9\pm0.09$ & $-1\pm0.14$ & $220$ & (2) \\
Z68-62 & 176.551 & 10.596 & 0.0102 & $9.1\pm0.05$ & $-2\pm0.2$ & $90$ & (2) \\
Z72-15 & 197.882 & 8.743 & 0.0133 & $9.5\pm0.08$ & $-1.7\pm0.16$ & $130$ & (2) \\
Z80-42 & 247.434 & 11.847 & 0.0172 & $9.4\pm0.09$ & $-1.2\pm0.1$ & - & (1) \\
Z97-38 & 174.811 & 14.992 & 0.0143 & $9.6\pm0.09$ & $-0.5\pm0.08$ & $180$ & (2) \\
\hline
LEDA1389480 & 162.297 & 11.149 & 0.0267 & $8.9\pm0.06$ & $-1.4\pm0.07$ & - & (1) \\
NGC2741 & 135.819 & 18.261 & 0.0116 & $9.2\pm0.13$ & $-0.7\pm0.08$ & $140$ & (2) \\
MRK802 & 210.943 & 14.868 & 0.0146 & $9.1\pm0.17$ & $-0.4\pm0.05$ & $170$ & (2) \\
LEDA213877 & 176.096 & 16.551 & 0.0129 & $9.7\pm0.1$ & $-3\pm0.9$ & - & (2) \\
MRK444 & 192.171 & 34.478 & 0.0142 & $9.4\pm0.14$ & $-0.3\pm0.07$ & $100$ & (2) \\
Z102-75 & 208.273 & 15.845 & 0.0102 & $10\pm0.09$ & $-2\pm0.3$ & - & (2) \\
NGC2607 & 128.486 & 26.973 & 0.0117 & $9.6\pm0.09$ & $-2\pm0.5$ & $50$ & (2) \\
MRK399 & 141.535 & 34.895 & 0.0164 & $9.5\pm0.15$ & $-0.4\pm0.12$ & $140$ & (2) \\
LEDA31379 & 158.914 & 28.566 & 0.0147 & $9.9\pm0.08$ & $-1\pm0.5$ & $150$ & (2) \\
MRK58 & 194.772 & 27.644 & 0.0181 & $9.7\pm0.08$ & $-0.4\pm0.08$ & $90$ & (2) \\
LEDA3998175 & 144.287 & 9.464 & 0.0225 & $8.8\pm0.06$ & $-2\pm0.2$ & - & (1) \\
Z65-27 & 156.287 & 13.601 & 0.0186 & $9.8\pm0.09$ & $-1\pm0.3$ & $50$ & (2) \\
UGC7017 & 180.594 & 29.862 & 0.0104 & $10\pm0.09$ & $-0.2\pm0.11$ & $290$ & (2) \\
LEDA86470 & 216.097 & 13.798 & 0.0168 & $9.4\pm0.1$ & $-0.8\pm0.09$ & $130$ & (2) \\
    \hline\\
    \end{tabular}
    \caption*{\textbf{Notes.} Columns five and six list total stellar mass and SFR measured in the fiber from the MPA-JHU catalog for SDSS DR8.
    Column seven lists the SDSS redshift of the target.
    Column eight lists the line width of the CO emission line used in this work.
    The source of the CO data (including FWHM) is indicated in the final column: (1) ALLSMOG \citep{ciconeFinalDataRelease2017} or (2) xCOLDGASS \citep[][no uncertainties on the measured FWHM of the CO lines provided]{saintongeXCOLDGASSComplete2017}.
    Galaxies in the upper section of the table were observed with NTT/EFOSC2 and those in the lower part with TNG/DOLORES.}
    \label{tab:target_info}
\end{table*}

\label{sec:tng_obs}
\begin{figure}[tbp]
    \centering
    \includegraphics[clip=true,trim=0.1cm 0.6cm 0.0cm 0.1cm, width=0.49\textwidth]{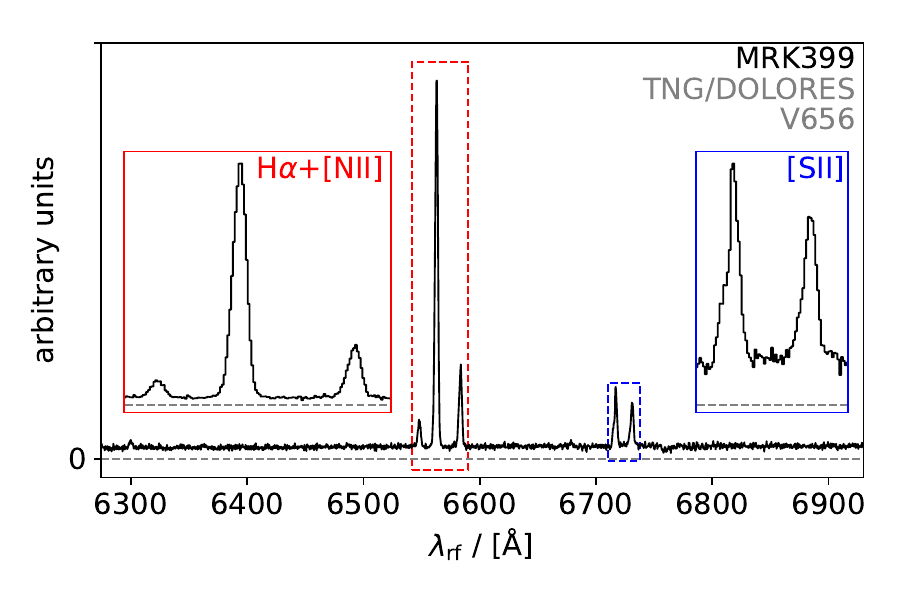}
\includegraphics[clip=true,trim=0.1cm 0.6cm 0.0cm 0.1cm, width=0.49\textwidth]{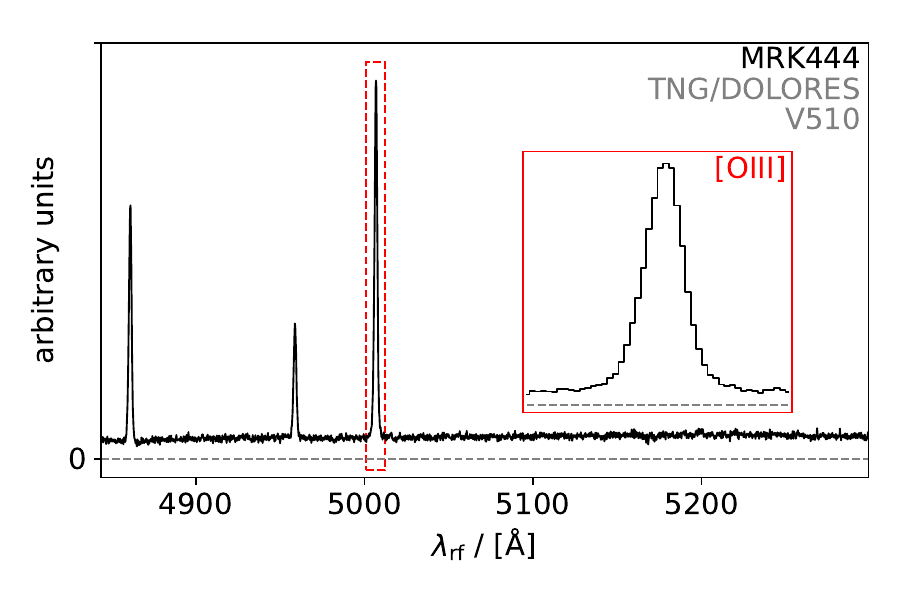}
    \caption{Example spectra obtained with TNG/DOLORES spectroscopy using the V656 and V510 grisms with a 1.0'' slit. We note that these spectra do not have their absolute flux calibrated, so units on the y-axis are arbitrary. The dashed gray line shows the zero flux level. Insets show a more detailed view of select emission lines, with the size of the zoomed-in region shown by the dashed colored boxes.}
    \label{fig:tng_examples}
\end{figure}

\subsection{TNG/DOLORES observations}
Of our sample, 14 targets were observed with TNG/DOLORES between 3 March and 5 March 2024 (Program ID: A48TAC\_16, PI: C. Vignali).
We obtained long-slit spectroscopy using the V510 and V656 grisms covering wavelength ranges of $4875$-$5325$\,\AA\ and $6232$-$6888$\,\AA\ respectively.
For the redshift range of our sample, this coverage includes respectively the [OIII]$\lambda5007$ (hereafter simply [OIII]) and H$\alpha$ emission lines.
With a 1.0'' slit and a $1\times1$ binning, this resulted in a nominal spectral resolution of $R\sim5950$ for the V510 setup and $R\sim5248$ for the V656 setup.
With each grism, we took a single exposure of between 800\,s and 3200\,s for each target, based on its SDSS fiber magnitudes in the r- and g-bands.
The slit was positioned on the peak of the galaxy's brightness profile in the acquisition image and its orientation was chosen to minimize atmospheric effects during the long exposures, and its alignment relative to the galaxy's geometry is therefore essentially random.
Examples of spectra with clearly detected emission lines are shown in Fig.~\ref{fig:tng_examples}.
Weather conditions were mixed, with seeing varying between 0.5'' and 2.0''.

\label{sec:ntt_obs}

\begin{figure}[tbp]
    \centering
    \includegraphics[clip=true,trim=0.1cm 0.6cm 0.0cm 0.1cm, width=0.49\textwidth]{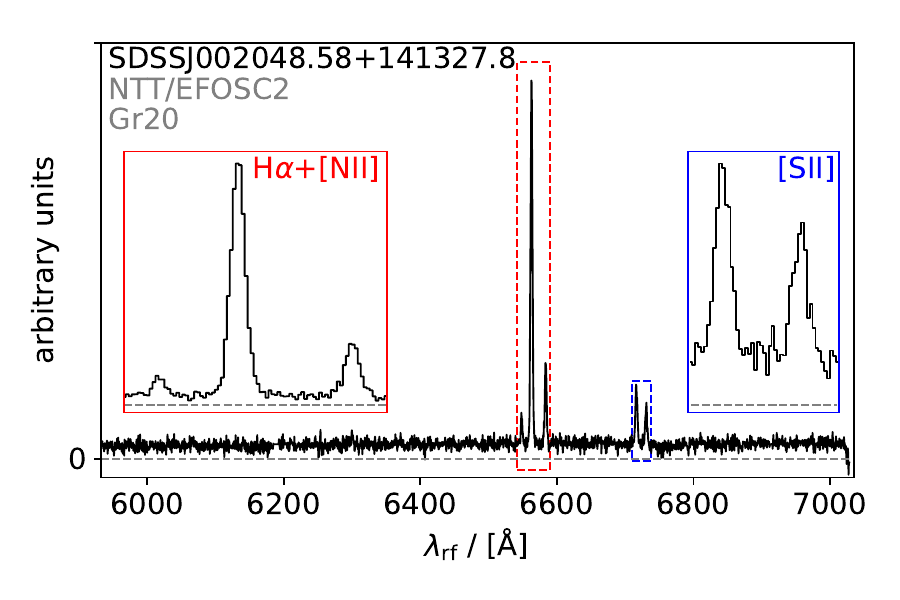}
    \includegraphics[clip=true,trim=0.1cm 0.6cm 0.0cm 0.1cm, width=0.49\textwidth]{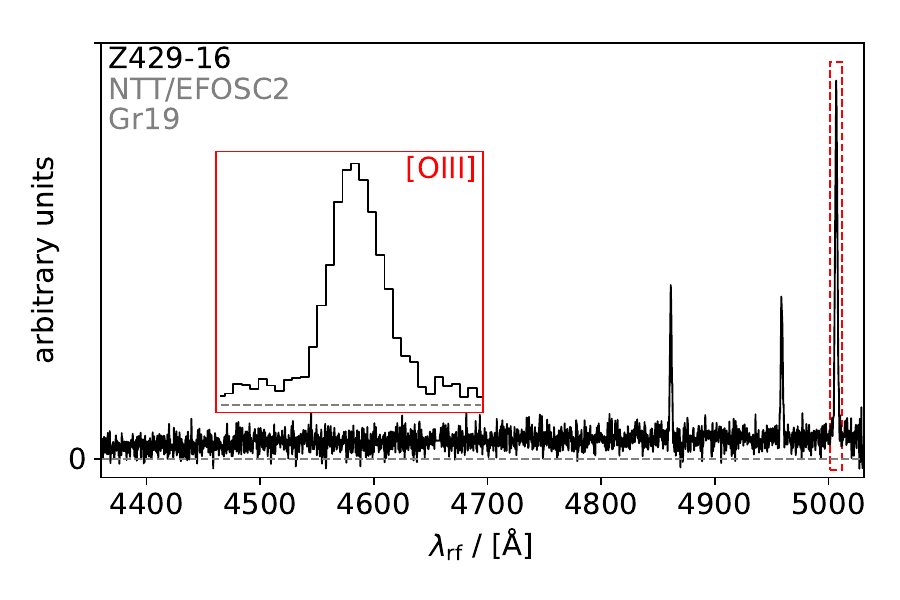}
    \caption{Example spectra obtained with NTT/EFOSC2 spectroscopy using  grisms 20 and 19 with a 0.5'' slit. We note that these spectra do not have their absolute flux calibrated, so units on the y-axis are arbitrary. The dashed gray line shows the zero flux level. Insets show a more detailed view of select emission lines, with the size of the zoomed-in region shown by the dashed colored boxes.}
    \label{fig:ntt_examples}
\end{figure}

\subsection{NTT/EFOSC2 observations}
The remaining 38 galaxies were observed with NTT/EFOSC2 (Program IDs: 113.269W, PI: B. Hagedorn).
For 25 we have coverage of both [OIII] using grism\#19 (gr19, covering 4441-5114\,\AA) and H$\alpha$ using grism\#20 (gr20, 6047-7147\,\AA), while 8 were observed only in [OIII] and 5 only in H$\alpha$.
With both grisms, a 0.5'' slit and a $1\times1$ binning were used leading to nominal spectral resolutions of $R\sim3338$ for gr19 and $R\sim3281$ for gr20.
Again, the slit was centered on the brightest part of the galaxy and its alignment is not matched to the galaxy's geometry.
With each grism, we took a single exposure of between 800\,s and 2810\,s for each target, based on its SDSS fiber magnitudes in the r- and g-bands.
For the additional, fainter targets mentioned in Section~\ref{sec:sample}, we took two exposures of 2810\,s each.
Examples of spectra with clearly detected emission lines are shown in Fig.~\ref{fig:ntt_examples}.
Weather conditions were mixed, with seeing varying between 1.0'' and 2.0''.

\subsection{CO(1-0) and CO(2-1) data}
\label{sec:co_data}
From the ALLSMOG and xCOLDGASS surveys, we obtain spectra of the CO(1-0) and CO(2-1) emission lines where available.
We use the FWHM reported in \citet{ciconeFinalDataRelease2017} and \citet{saintongeXCOLDGASSComplete2017} in our analysis. 
The native spectral resolution of the heterodyne instruments used for these observations is extremely high ($R\sim10^5-10^6$, although most spectra are rebinned to increase the S/N), hence CO line width measurements are not limited by spectral resolution.

Out of 52 galaxies in our sample 35 are detected in CO emission at the $3\sigma$ level and 17 are non-detections.
The CO emission line spectra used in this work are shown alongside the optical emission line spectra in Appendices \ref{app:ntt_specs} and \ref{app:tng_specs}.

\subsection{Data reduction}
\label{sec:data_redux}
We use version 1.17.4 of \texttt{PypeIt}\footnote{\url{https://pypeit.readthedocs.io/en/latest/}} \citep{pypeit:joss_pub,pypeit:zenodo}, a Python package for semi-automated reduction of astronomical slit-based spectroscopy, to reduce both TNG/DOLORES and NTT/EFOSC2 spectroscopy.
\texttt{PypeIt} has specific default parameter configurations for both EFOSC2 and DOLORES, which we adopt for the reduction process.

Since our analysis focuses on emission lines rather than continuum emission, we use a boxcar extraction profile with a radius of 1.5'' for the extraction of 1D spectra, to ensure we do not exclude emission line flux in cases where its spatial distribution deviates from that of the continuum.
For observations with EFOSC2's gr19, no strong sky lines fall into the observed wavelength range, so we do not perform a correction for flexure in the spectral direction.

For wavelength calibration and to determine the spectral resolution, we used the 1D spectra of the arc lamp calibration exposures taken at the beginning of each night.
We find a median spectral resolution (FWHM) of 78\,km/s (52\,km/s) for \ha with EFOSC2/gr20 (DOLORES/V656) and 114\,km/s (48\,km/s) for [OIII] with EFOSC2/gr19 (DOLORES/V510).
The spectral resolution for [OIII] with EFOSC2/gr19 is comparatively poor because the line is close to the limit of the spectral range.

\section{Analysis}
\label{sec:methods}

\subsection{Emission line fitting}
\label{sec:methods:line_fitting}
For a first assessment of the signal-to-noise ratio (S/N, which we define as the amplitude of the emission line divided by the rms of the spectrum estimated from line-free regions) of the targeted emission lines, and to estimate their width, we perform an initial fit to the spectra in regions around the \ha (6535-6600\,\AA) and [OIII] (4985-5025\,\AA) emission lines.
We subtract a constant pseudo-continuum\footnote{We do not fit a more complex continuum model in this step, due to the absence of strong absorption line features in the data necessary to adequately constrain such a fit.
While the overlap between the stellar \ha absorption and nebular \ha emission line features could potentially affect the resulting measured line profiles, the absence of strong absorption line features and the strength of the emission lines in most of our spectra mean this is unlikely to play an important role here.} from the 1D spectra extracted from NTT/EFOSC2 and TNG/DOLORES spectroscopy around the \ha and [OIII] emission lines, and use the standard deviation of the flux density in line free parts of the spectra to estimate the noise level.

We use the \texttt{lmfit} python package \citep{newville2014lmfit} to fit both the \ha and [OIII] lines with a single Gaussian function, in a range corresponding to a Doppler shift of $\pm$350\,km/s from the expected line center based on the redshift retrieved from the SDSS data.
In case of the \ha line, the [NII] doublet is also modeled with a set of two Gaussian functions with shared kinematics (central velocity and velocity dispersion) and with relative amplitudes fixed to a ratio of 0.34.
The Gaussian emission line templates are convolved with the instrumental resolution of EFOSC2 or DOLORES, as applicable, before fitting.
Based on this initial fit, we consider an emission line detected if it has S/N$>$3.
Out of 43 galaxies observed in H$\alpha$, 39 are detected, and for [OIII] we detect 26 out of 46 observed galaxies.
The line widths obtained from the initial fit are listed in Table~\ref{tab:highres_info}.

\subsection{Fitting of line wings}
\label{sec:methods:outflow_fitting}
To account for the presence of potential outflow signatures in the high-resolution spectroscopy, we iteratively fit the \ha and [OIII] emission lines with a more flexible model once the line center and overall dispersion have been determined.
This model uses up to three independent Gaussian components that are fitted simultaneously to the continuum-subtracted line spectrum. 
This enables us to to model both blue- and red-shifted wings, which may not be symmetric in the presence of dust extinction from the disk or from the outflow itself \citep[see e.g.,][]{Hagedorn2026MassiveMultiphase}.
For each Gaussian component velocity dispersion, $\sigma$, is constrained to be greater than the instrumental resolution at the wavelength of the observation.
Other parameters are left free to vary within the spectral range of the fit ($\pm350$\,km/s).
This range was chosen after a visual inspection of the spectra to be broad enough to include any high-velocity features but narrow enough to exclude neighboring lines, such as the [NII] doublet around H$\alpha$.
We fit in a three-step iterative process (with one, two, and three Gaussian components respectively) using the \texttt{PySpecKit} package \citep{ginsburgPyspeckitSpectroscopicAnalysis2022, pyspeckitSoftware} with initial parameters informed by the residuals of the fit in the previous iteration.
Thanks to the high resolution of the spectra and the narrow emission lines in the target galaxies, there is no blending between the \ha line and the nearby [NII] doublet (see examples in Figures \ref{fig:tng_examples} and \ref{fig:ntt_examples}), allowing us to fit the \ha line in isolation in this step.

The best-fit model is chosen from among the three iterations based on the Akaike information criterion \citep[AIC,][]{Akaike1974}.
Each Gaussian component contributes three free parameters to the model, and the maximum likelihood is estimated from the residual sum of squares of the range of the fit, which we use to determine the number of independent data points.

We do not immediately assume that a best-fit model with more than one component implies the presence of outflows, as these components may trace other mechanisms, such as tidal structures, unresolved merging companions, or inflows.
Instead, we use the distribution of central velocities and velocity dispersions to identify likely outflow signatures as described in Section~\ref{sec:res:outflows}.

\section{Results}
\label{sec:results}

\subsection{Comparison between high-resolution \ha and CO line widths}
\label{sec:res:ha_co_comp}

\begin{figure}[]
    \centering
    \includegraphics[clip=true, trim=0.6cm 0.3cm 0.4cm 0.2cm,width=0.49\textwidth]{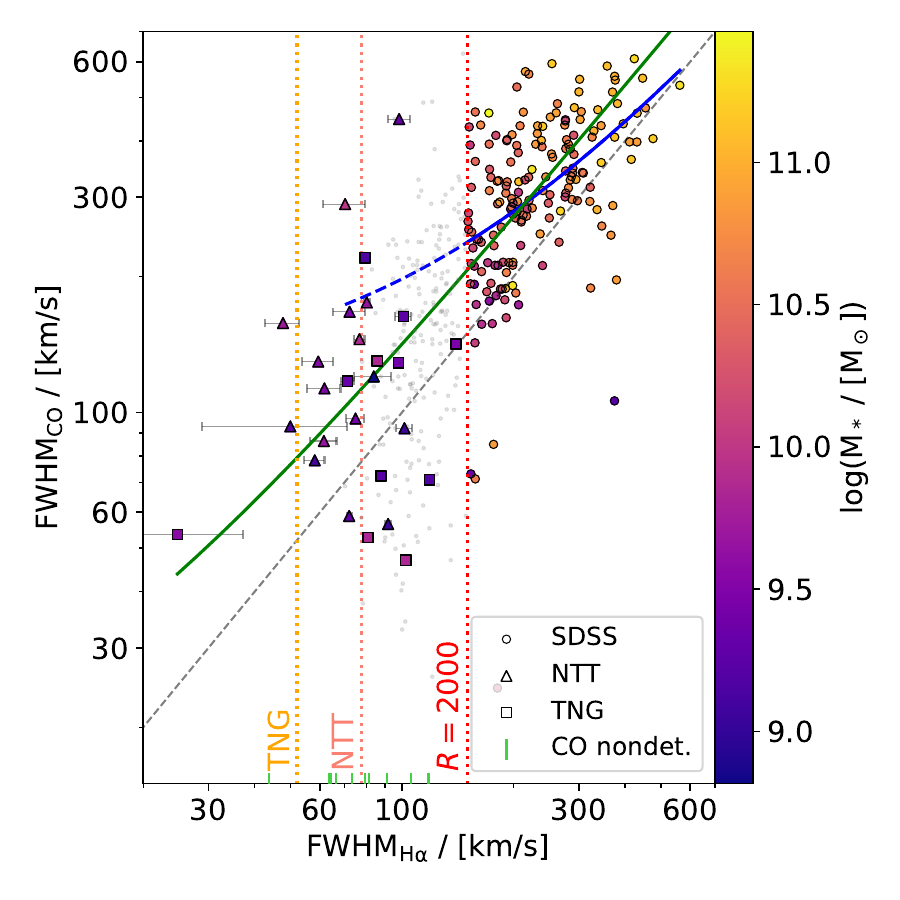}
    \caption{Same as Fig.~\ref{fig:allgass_ha_co_fwhm}, but with line widths inferred from high resolution spectroscopy replacing data unresolved by SDSS (the unresolved data are still shown as gray dots).
    The solid green line shows a linear fit to the combined SDSS resolved and high-resolution data. 
    The blue solid line shows the same linear fit to the resolved SDSS data as in Fig.~\ref{fig:allgass_ha_co_fwhm}, extrapolated to the SDSS-unresolved regime as a dashed line.
    Targets without detections in CO are marked with vertical line markers on the x-axis at the position corresponding to their \ha line widths inferred from their NTT or TNG spectra.
    Dotted vertical lines show the approximate SDSS resolution $R=2000$ (FWHM $\sim150$\,km/s), and the median resolution for the sample achieved with TNG (FWHM $\sim52$\,km/s) and NTT (FWHM $\sim78$\,km/s) for the H$\alpha$ line.}
    \label{fig:co_ha_fwhm}
\end{figure}

We use \ha line widths inferred from the initial single-Gaussian fit to the new high-resolution spectra to test the effects of an improved \ha spectral resolution on the relation between \ha and CO line widths previously shown in Fig.~\ref{fig:allgass_ha_co_fwhm}.
The results are shown in Fig.~\ref{fig:co_ha_fwhm}, which includes the original SDSS data points for a comparison (gray dots). 
The \ha FWHM values measured with the higher resolution data are, as expected, smaller than the previously measured SDSS values.
As a result, the correlation between CO and \ha line widths holds down to the lowest FWHM values. The downturn that was visible in Fig.~\ref{fig:allgass_ha_co_fwhm} has disappeared, confirming that it was indeed due to the insufficient SDSS resolution making an accurate determination of \ha~line widths impossible in this regime.

Moreover, thanks to the new higher resolution data, we can confirm the trend of broader CO than H$\alpha$ lines  even in the low M$_*$ regime. 
Differences in aperture size and placement between optical observations and SDSS ($3''$ fiber), TNG ($1''$), NTT ($0.5''$) and the CO observations with APEX ($27''$ at 1.3~mm) and IRAM ($22''$ at 3~mm) are likely a major factor in this systematic difference.
In Appendix~\ref{app:linewidths}, where we compare \ha line widths inferred from SDSS with the new values computed from the higher-resolution spectra, we show that such aperture effects are unlikely to play a significant role in the observed differences in optical line widths between SDSS on the one hand, and TNG and NTT on the other hand, however.

\begin{table*}
\caption{\raggedright Results from the analysis of high-resolution spectroscopy of the sample.}
    \centering
    \begin{tabular}{lcccccc}
    \hline
    \hline
Target & S/N$_\mathrm{H\alpha}$ & S/N$_\mathrm{[OIII]}$ & FWHM$_\mathrm{H\alpha}$ & FWHM$_\mathrm{[OIII]}$ & $f_\mathrm{out,H\alpha}$ & $f_\mathrm{out,[OIII]}$ \\
 & & & [km/s] & [km/s] & & \\
\hline
2MASSJ00114318+1428010 & 107 & 11 & $92\pm1.5$ & $90\pm11$ & $0.2\pm0.09$ & - \\
AGC244381 & 10 & - & $70\pm14$ & - & - & - \\
IC159 & 36 & 7 & $50\pm5$ & $100\pm20$ & - & - \\
IC5292 & 59 & - & $60\pm3$ & - & - & - \\
LEDA1278497 & 22 & 4 & $110\pm6$ & $100\pm40$ & - & - \\
LEDA1397680 & 82 & - & $91\pm1.2$ & - & - & - \\
LEDA1417541 & - & - & - & - & - & - \\
LEDA1439867 & - & 15 & - & $70\pm6$ & - & - \\
LEDA1446233 & - & - & - & - & - & - \\
LEDA1452135 & - & - & - & - & - & - \\
LEDA1464874 & 7 & - & $70\pm17$ & - & - & - \\
LEDA3091160 & 44 & 8 & $100\pm7$ & $80\pm18$ & - & - \\
LEDA3127469 & 39 & - & $40\pm5$ & - & - & - \\
LEDA3530 & 164 & 23 & $72\pm1.1$ & $90\pm3$ & $0.1\pm0.04$ & - \\
LEDA49155 & - & 8 & - & $100\pm17$ & - & - \\
MCG+00-34-035 & - & - & - & - & - & - \\
NGC7401 & 13 & - & $80\pm10$ & - & - & - \\
NGC99 & 30 & - & $60\pm5$ & - & - & - \\
SDSSJ002048.58+141327.8 & 86 & 13 & $118\pm1.5$ & $120\pm5$ & - & - \\
SDSSJ011437.36+011053.4 & - & - & - & - & - & - \\
SDSSJ120108.03+152353.6 & - & - & - & - & - & - \\
SDSSJ135906.39+175604.9 & - & 11 & - & $110\pm11$ & - & - \\
SDSSJ140317.75+100340.3 & 81 & 25 & $60\pm4$ & $70\pm2$ & - & - \\
UGC10005 & - & - & - & - & - & - \\
UGC11992 & - & - & - & - & - & - \\
UGC272 & 7 & - & $<60$ & - & - & - \\
UGC6838 & 25 & - & $100\pm5$ & - & - & - \\
UGC9359 & - & - & - & - & - & - \\
Z38-48 & 37 & 4 & $80\pm3$ & $100\pm40$ & - & - \\
Z383-30 & 25 & - & $60\pm6$ & - & - & - \\
Z386-19 & 25 & - & $60\pm5$ & - & - & - \\
Z429-16 & 90 & 22 & $80\pm1.5$ & $90\pm4$ & $0.3\pm0.11$ & - \\
Z429-21 & 50 & 6 & $80\pm2$ & $<110$ & - & - \\
Z436-34 & 22 & - & $60\pm6$ & - & - & - \\
Z68-62 & 17 & 5 & $70\pm7$ & $100\pm40$ & - & - \\
Z72-15 & 14 & - & $70\pm9$ & - & - & - \\
Z80-42 & 35 & 6 & $70\pm4$ & $80\pm16$ & - & - \\
Z97-38 & - & - & - & - & - & - \\
\hline
LEDA1389480 & 41 & 11 & $70\pm3$ & $70\pm5$ & - & - \\
NGC2741 & 530 & - & $81\pm1.1$ & - & - & - \\
MRK802 & 720 & - & $79\pm0.6$ & - & $0.31\pm0.015$ & - \\
LEDA213877 & - & - & - & - & - & - \\
MRK444 & 510 & 40 & $102\pm0.7$ & $103\pm0.8$ & $0.12\pm0.019$ & $<0.5$ \\
Z102-75 & 24 & - & $100\pm3$ & - & - & - \\
NGC2607 & 4 & 4 & $20\pm13$ & $50\pm20$ & - & - \\
MRK399 & 229 & 26 & $118\pm0.8$ & $122\pm1.3$ & - & $<0.4$ \\
LEDA31379 & 17 & - & $100\pm5$ & - & $0.4\pm0.12$ & - \\
MRK58 & 159 & 4 & $88\pm0.9$ & $130\pm19$ & $0.2\pm0.04$ & - \\
LEDA3998175 & - & - & - & - & - & - \\
Z65-27 & 88 & 32 & $80\pm1.6$ & $114\pm1.7$ & $0.4\pm0.06$ & $0.4\pm0.16$ \\
UGC7017 & 104 & 5 & $140\pm1.5$ & $110\pm17$ & $0.2\pm0.08$ & - \\
LEDA86470 & 160 & 17 & $86\pm0.7$ & $110\pm3$ & $0.2\pm0.07$ & - \\
    \hline\\
    \end{tabular}
    \caption*{\textbf{Notes.} Columns two and three list the emission line FWHM (deconvolved from instrumental resolution) for \ha and [OIII] derived from the single-component fit to the high-resolution spectroscopy where the fit was successful.
    Columns four and five show the ratio of flux in outflow-like fit components to the total flux in the emission line derived from the multi-component fit to the high-resolution spectroscopy for galaxies in which the best-fit model has outflow-like components (see Section \ref{sec:res:outflows}).
    Galaxies in the upper section of the table were observed with NTT/EFOSC2 and those in the lower part with TNG/DOLORES.}
    \label{tab:highres_info}
\end{table*}

\subsection{\ha line wings and asymmetries}
\label{sec:res:asymmetry_wing_tests}
For galaxies in which the \ha emission line is detected at sufficiently high S/N, we search for the presence of high-velocity wings or asymmetries.
Such features can indicate galaxy-scale outflows that may have remained undetected in the unresolved SDSS spectra.

We compute the excess kurtosis (hereafter simply kurtosis) of the \ha line profile in the sample to assess the presence of high-velocity wings.
The kurtosis of a distribution, or fourth standardized moment, measures its ``tailed-ness'', so for emission lines with broad wings due to outflows we expect an excess kurtosis relative to a Gaussian line profile.
Measures of emission line profile kurtosis have previously been used to identify galactic outflows in integrated spectroscopy \citep{yuNonGaussianityOpticalEmission2022}.
To ensure that the kurtosis measurement is not dominated by noise, we calculate the kurtosis using the spectral range where the emission line exceeds 3 times the rms noise level of the spectrum.
We calculate the kurtosis in this way for galaxies which are detected at S/N$>$10 in H$\alpha$, to exclude galaxies for which this cut would remove significant amounts of flux, possibly distorting the line profile.
This leaves us with a sub-sample of 33 galaxies.

\begin{figure}
    \centering
    \includegraphics[clip=true, trim=0.6cm 0.3cm 0.4cm 0.5cm, width=0.4\textwidth]{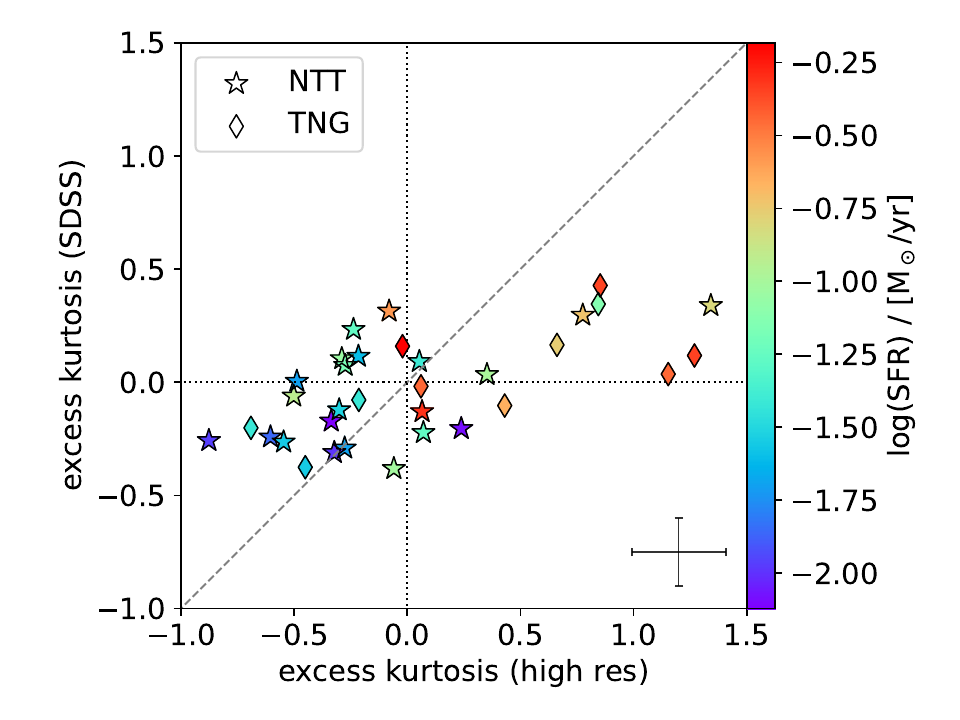}
    \caption{Comparison between line profile kurtosis of SDSS spectra and corresponding high-resolution observations. The color scale corresponds to the SFR of each galaxy measured in the SDSS fiber. The error bars int he bottom right show the typical statistical uncertainty on the calculated values.}
    \label{fig:kurtosis_comp}
\end{figure}

Fig.~\ref{fig:kurtosis_comp} shows the \ha line profile kurtosis for each galaxy calculated from SDSS spectroscopy vs that calculated from the high-resolution NTT or TNG data.
The SDSS spectroscopy produces line profiles with kurtosis values between $\pm0.5$, whereas the values for the high-resolution spectra are scattered more broadly, with a few notable examples exhibiting kurtosis values $>$1.
This is due to the lines being unresolved in the SDSS spectroscopy, resulting in line shapes close to the, approximately Gaussian, instrumental profile and corresponding expected kurtosis of zero.
Galaxies with above average SFR (median $\sfr=10^{-1}\,\mathrm{\msun/yr}$ for galaxies in the sample with kurtosis measurements) tend to also exhibit leptokurtic line profiles (i.e., kurtosis $>$0), as expected if the excess kurtosis is due to wing emission from high-velocity gas in star-formation driven outflows.

We perform a similar test using non-parametric percentile velocities $v02$, $v50$, and $v98$ which correspond to the range Doppler velocities in which 2.3\%, 50\%, and 97.7\% of the total emission line flux are contained.
We use the difference between the absolute values of $v02$ and $v98$ to assess asymmetries in the spectra.
Blue asymmetries (i.e., $|v02|-|v98|>0$) are usually interpreted as signatures of outflows \citep[e.g.,][]{ciconeOutflowsComplexStellar2016}, where emission from red-shifted outflowing gas is attenuated by dust in the galaxy disk.
We adjust both velocities by the central velocity of the line $v50$, in order to account for possible differences in line centroids which may be due to either aperture (i.e., SDSS fiber vs slit of the NTT/TNG spectra) or to slit positioning effects, especially for targets that are very extended.
On average, we find a difference of $\sim15\,$km/s in $v50$ between SDSS and higher-resolution spectra.
One extreme example of this is the spectrum of UGC7017, for which there is a significant velocity shift in the lines observed with SDSS spectroscopy vs those in the high-resolution spectra (see Fig.~\ref{fig:ha_cards_tng_2}).
UGC7017 is much more extended than the 3'' SDSS fiber, making it susceptible to aperture and targeting effects.

\begin{figure}
    \centering
    \includegraphics[clip=true, trim=0.6cm 0.3cm 0.4cm 0.5cm, width=0.4\textwidth]{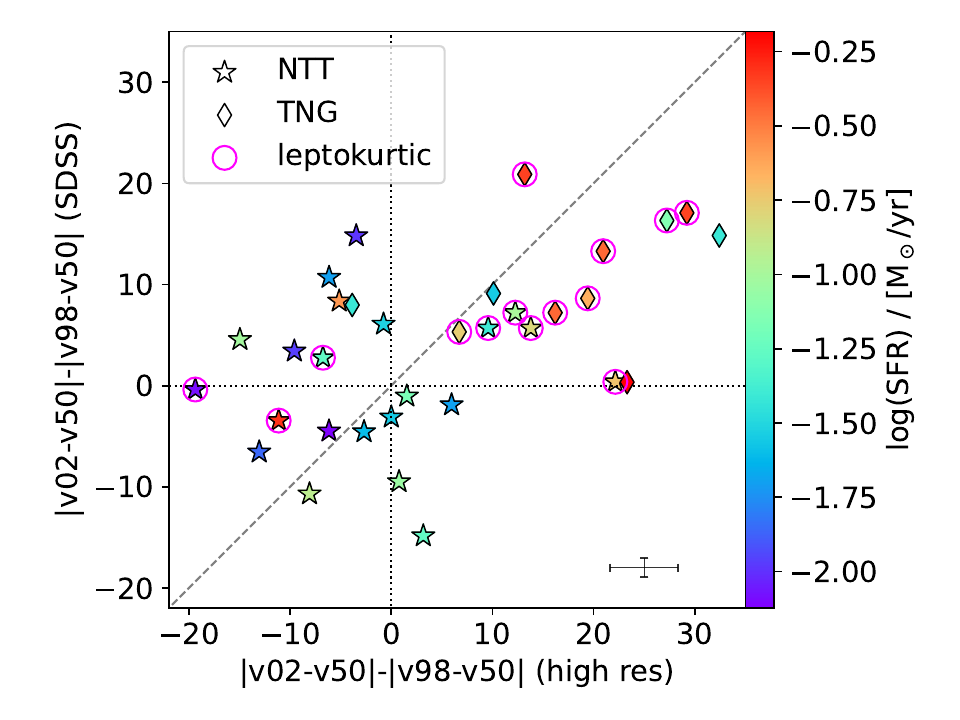}
    \caption{Comparison between absolute percentile velocity differences. The color scale corresponds to the SFR of each galaxy measured in the SDSS fiber. Magenta circles mark galaxies with leptokurtic \ha line profiles (i.e., kurtosis$>$0). The error bars int he bottom right show the typical statistical uncertainty on the calculated values.}
    \label{fig:skew95_comp}
\end{figure}

After this adjustment, the difference $|v02-v50|-|v98-v50|$ should primarily capture asymmetries in the line profile, with positive values corresponding to excess flux on the blue-shifted side of the profile, and negative values to excess flux on the red-shifted side.
From Fig.~\ref{fig:skew95_comp} it is evident that the high-resolution data cover a larger range of values in this parameter, with a few notable examples exceeding 20\,km/s.
Most of the galaxies exhibiting strong blue asymmetries in \ha also have leptokurtic line profiles (see magenta circles in Fig.~\ref{fig:skew95_comp}).
We thus confirm the presence, in several targets, of line asymmetries compatible with galactic outflows, which remain undetected in the lower resolution SDSS spectra.

\subsection{Comparison of [OIII]5007 and \ha line profiles}
\label{sec:res:o3_comp}
In the previous sections, we focused on the \ha emission line as a tracer of ionized gas kinematics, since we have more and higher S/N detections of this line in our sample than of the [OIII] line.
In the following, we will compare the \ha and [OIII] lines in galaxies in our high-resolution sample where they are both detected, before moving on to the identification and characterization of outflows in the data.

\begin{figure}
    \includegraphics[clip=true, trim=0.6cm 0.3cm 0.4cm 0.5cm, width=0.4\textwidth]{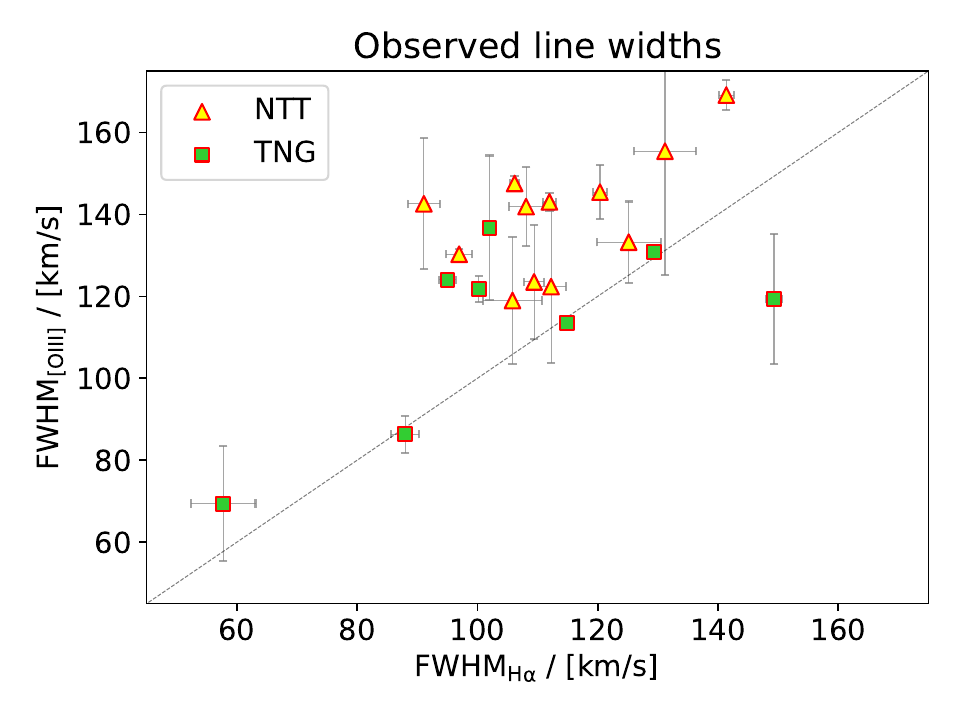}
    \includegraphics[clip=true, trim=0.6cm 0.3cm 0.4cm 0.5cm, width=0.4\textwidth]{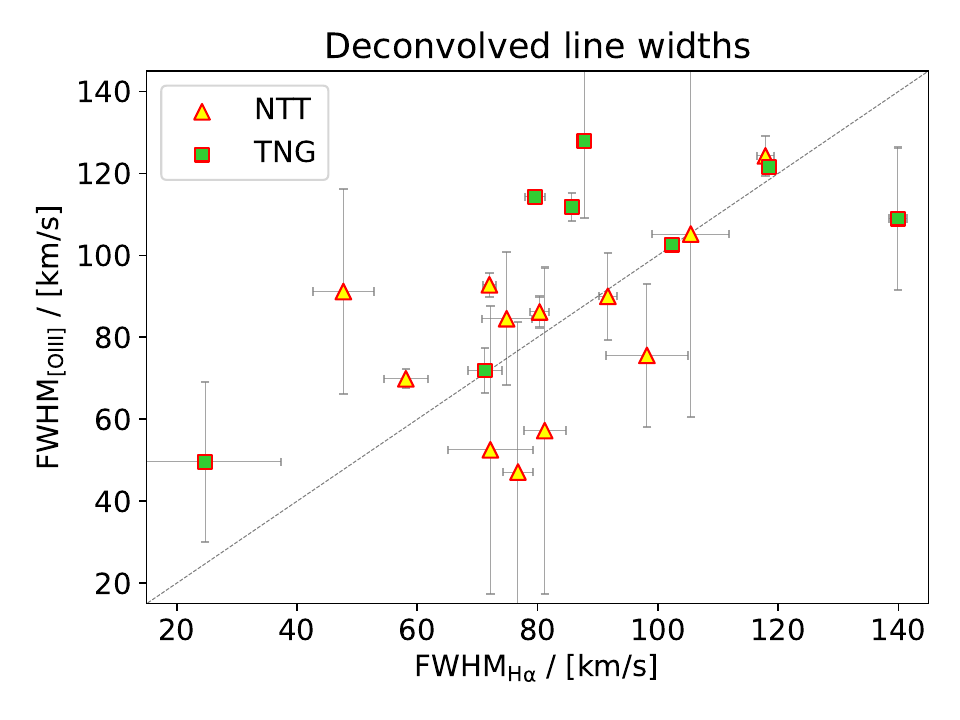}
    \caption{Comparison between H$\alpha$ and [OIII]5007 line widths observed with high-resolution spectroscopy (top) and deconvolved from instrumental resolution (bottom).}
    \label{fig:ha_o3_fwhm_comp}
\end{figure}

The observed [OIII] lines are generally broader than the \ha lines in these galaxies, as can be seen in the top panel of Fig.~\ref{fig:ha_o3_fwhm_comp}.
However, this trend appears to be mainly driven by observations with NTT/EFOSC2.
In these, the [OIII] line falls close to the end of the wavelength range covered by the grism used (gr19), which results in a relatively poor spectral resolution ($\sim100$\,km/s).
After subtracting the instrumental dispersion in quadrature for both lines (bottom panel in Fig.~\ref{fig:ha_o3_fwhm_comp}), the data are scattered around the one-to-one relation between the two lines.
We caution that line widths derived from single-component Gaussian fits may not accurately characterize a well resolved line profile, especially if there are asymmetries or broad wings present.
In Section \ref{sec:res:outflows}, we compare the two lines in greater detail with a fit function better suited to model any additional components detected in the line profiles.

\subsection{Outflow characterization}
\label{sec:res:outflows}
To account for the asymmetries and non-Gaussianity of line profiles observed in the new high-resolution spectra, we perform a fit with the multi-component model described in Section~\ref{sec:methods:outflow_fitting}.
Plots with the best-fit models are shown in Appendices \ref{app:ntt_specs} and \ref{app:tng_specs} and the results are shown in Tables~\ref{tab:highres_info} and \ref{tab:outflow_props}.
The best-fit model requires a single Gaussian component for 24, two components for 12, and three components for 3 out of 39 galaxies with \ha detections.
For the 26 [OIII] detected galaxies, these numbers are 23, 2, and 1 respectively.
Using the same fitting routine for the SDSS data shows that all line profiles are best described with a single Gaussian.

\begin{figure}[tbp]
    \centering
    \includegraphics[clip=true, trim=0.6cm 0.3cm 0.4cm 0.5cm, width=0.4\textwidth]{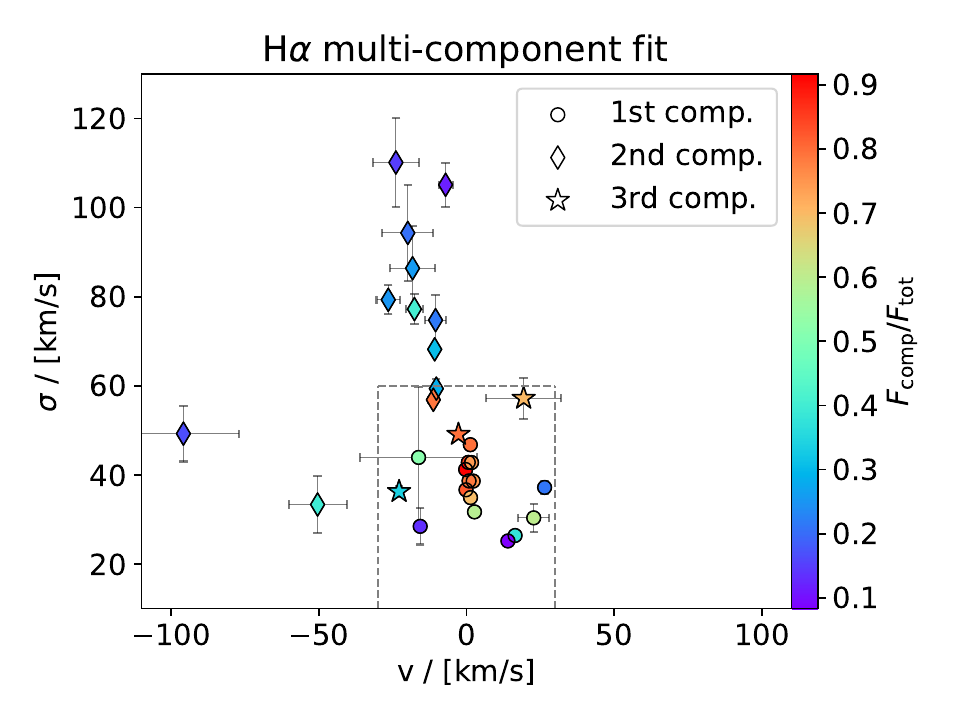}
    \includegraphics[clip=true, trim=0.6cm 0.3cm 0.4cm 0.5cm, width=0.4\textwidth]{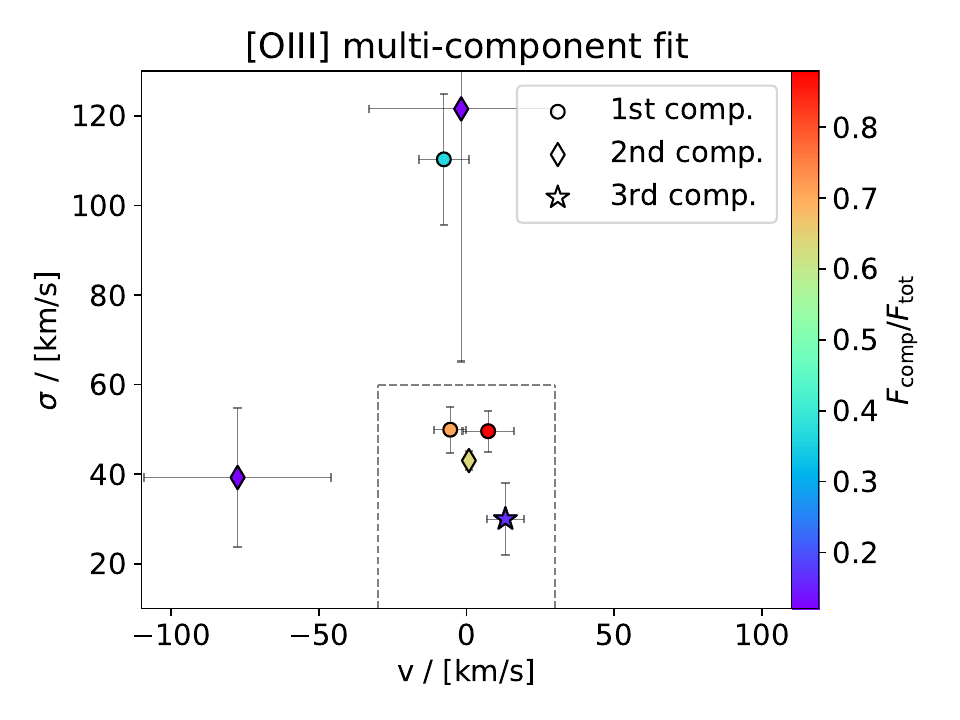}
    \caption{Central velocity $v$ and velocity dispersion $\sigma$ of all components in the best fit multi-Gaussian model across all galaxies in which the relevant line is detected and required multiple components to fit. The top panel shows the results of the fit to the \ha emission line and the bottom panel the [OIII] emission line. The color scale shows the flux of the component relative to the total flux summed over all components in the best-fit model. The dashed gray rectangle corresponds to $\pm30$\,km/s in $v$ and 60\,km/s in $\sigma$, and marks the separation between components tracing potential outflow signatures and those tracing rotating gas. Numbering of the components only reflects the order in which they are added during the iterative fitting procedure.}
    \label{fig:multi_comp_vel_sigma}
\end{figure}

Fig.~\ref{fig:multi_comp_vel_sigma} shows the central velocity $v$ and dispersion $\sigma$ of the components from the best-fit models across all 15 (3) galaxies in which the \ha ([OIII]) line is detected and required more than one component to fit.
We divide between narrow components ($\sigma<60$\,km/s) with small velocity offsets ($|v|<30$\,km/s), and broad or blue-shifted components, that exceed either of the criteria.
We classify the former as tracing quiescent ISM, while the latter trace potential outflow signatures.
The limits for this classification scheme are chosen such that line components which account for a majority of the total flux of the emission line ($F_\mathrm{comp}/F_\mathrm{tot}>0.5$) all lie within the bounds we adopt for quiescent components, as shown by the color scale in Fig~\ref{fig:multi_comp_vel_sigma}, and components that lie outside of the chosen boundaries are all either broad or blue-shifted but not red-shifted, which is consistent with the expected signatures of outflowing high-velocity gas.
In cases where both blue- and red-shifted outflowing high-velocity gas is observed, this leads to relatively symmetric broad wings in the emission line profile, corresponding to the components in the top half of the plots in Fig.~\ref{fig:multi_comp_vel_sigma}.
Where the red-shifted side of the outflow is attenuated by extinction from dust in the galaxy disk and only the blue-shifted part of the outflow is observed, line profiles show a blue asymmetry, corresponding to components on the left in Fig.~\ref{fig:multi_comp_vel_sigma}.

In the absence of additional constraints on the kinematics of the quiescent ISM (e.g., from stellar kinematics), the accuracy of this classification scheme is uncertain.
We do however, find good agreement of the scheme with the non-parametric measures described in Section \ref{sec:res:asymmetry_wing_tests}. 
Components outside the adopted bounds are all associated with lines that have either relatively high kurtosis ($>0.35$) or blue excess ($|v02-v50|-|v98-v50|>20$\,km/s).

Out of 39 galaxies with an \ha detection in our sample, 10 show potential outflow signatures (26\%), when applying the above criteria.
For [OIII] this is true for only 3 out of 26 (12\%).
However, if we consider only galaxies with emission line detections at S/N$>$10, we observe potential outflow signatures in 10/33 or 30\% of galaxies in \ha and 3/12 (26\%) in [OIII].
Two effects may play a role in this.
On the one hand, typically faint outflow-like components are more likely to go undetected in low S/N spectra, and on the other hand, stronger emission lines correlate with higher SFR, which is linked to outflow incidence.

Galaxies with a high S/N$>$10 detection of at least one of the \ha and [OIII] lines, and in which an outflow is detected in at least one of these two tracers have a median star formation rate of $\log(\mathrm{SFR/[M_\odot yr^{-1}]})=-0.72$ and a median stellar mass of $\log(M_*/\mathrm{[M_\odot]})=9.38$ compared to $\log(\mathrm{SFR/[M_\odot yr^{-1}]})=-1.41$ and $\log(M_*/\mathrm{[M_\odot]})=9.31$ for galaxies with S/N$>$10 in at least one tracer but no outflow detection in either.
So, even when correcting for S/N effects, the incidence of outflows is higher for higher SFRs, as is to be expected if feedback from star formation is the primary driver for these outflows.

\section{Discussion}
\label{sec:discussion}
In this section we estimate physical properties of the observed outflows and use them to infer mass outflow rates and mass loading factors.
We also discuss the results of our outflow analysis in the context of previous observations of, and theoretical predictions for, the outflow incidence in the low-$M_*$ regime.
Before we do so, it is important to point out the limitations of integrated spectroscopy for the purpose of outflow identification.
While the broad and asymmetric components we identify in the observed emission line profiles are consistent with signatures expected from outflowing gas, they could, in principle, also be attributed to inflowing gas, merging companions, components of tidal origin, or even rotation.
Without knowledge of where the emitting gas is located relative to the galaxy and the observer, it is not possible to conclusively break this degeneracy.

It is, however, reassuring that the potential outflow signatures we identify are all blue-shifted relative to their quiescent counterparts (see Fig.~\ref{fig:multi_comp_vel_sigma}).
This is consistent with extinction from the ISM affecting signatures of red-shifted outflow components, which must be on the far side of the galaxy disk from the observer.
For inflows, the effect of extinction, if present, would be reversed and we would expect to see more red-shifted high-velocity components, if there was a significant contribution from inflowing gas in our sample.
Nonetheless, it is possible that our analysis overestimates the real outflow incidence rate by including galaxies in which inflows, mergers, or similar events produce kinematic signatures indistinguishable from outflows using integrated spectroscopy.

We also note that our sample is not representative of the low-mass galaxy population in general, due to selection and sensitivity effects.
The outflow incidence we find in the sample therefore is not a measure of the incidence in the overall population of low-mass galaxies.

\begin{table*}
\caption{\raggedright Outflow properties}
    \centering
    \begin{tabular}{lcccccc}
    \hline
    \hline
Target/sub-sample & $|v_\mathrm{out}|$ & $r_{50}$ & $\dot{M}_\mathrm{out}$ & $\eta$ & incidence \\
 & [km/s] & [kpc] & $[\mathrm{M_\odot/kyr}]$ & & \\
\hline
2MASSJ00114318+1428010 & $120\pm4$ & $0.45\pm0.003$ & $23\pm19$ & $\sim0.2$ & \\
LEDA3530 & $120\pm3$ & $0.91\pm0.006$ & $12\pm10$ & $\sim0.07$ & \\
Z429-16 & $130\pm5$ & $1.1\pm0.06$ & $19\pm15$ & $\sim0.1$ & \\
MRK802 & $111\pm1.9$ & $1.03\pm0.004$ & $52\pm18$ & $\sim0.14$ & \\
MRK444 & $120\pm2$ & $1\pm0.01$ & $10\pm2$ & $\sim0.02$ & \\
LEDA31379 & $110\pm15$ & $1.77\pm0.002$ & $0.1\pm0.09$ & $\sim0.003$ & \\
MRK58 & $120\pm3$ & $2.48\pm0.008$ & $8\pm3$ & $\sim0.02$ & \\
Z65-27 & $130\pm5$ & $1.55\pm0.01$ & $1.4\pm0.5$ & $\sim0.02$ & \\
UGC7017 & $150\pm4$ & $1.9\pm0.09$ & $\sim30$ & $\sim0.05$ & \\
LEDA86470 & $100\pm2$ & $0.86\pm0.006$ & $6\pm2$ & $\sim0.03$ & \\
\hline
full sample & $120\pm6$ & $1.3\pm0.03$ & $\sim16$ & $\sim0.07$ & 30\% \\
\hline
logSFR $<$ -0.74 & $120\pm7$ & $1.11\pm0.006$ & $\sim8$ & $\sim0.07$ & 20\% \\
logSFR $>$ -0.74 & $130\pm3$ & $1.5\pm0.05$ & $\sim20$ & $\sim0.07$ & 60\% \\
\hline
log$M_*<$ 9.40 & $120\pm3$ & $0.9\pm0.03$ & $22\pm14$ & $\sim0.11$ & 30\%  \\
log$M_*>$ 9.40 & $130\pm7$ & $1.7\pm0.04$ & $\sim10$ & $\sim0.02$ & 40\%  \\
\hline
$f_\mathrm{mol}<$ 0.16 & $120\pm8$ & $1.66\pm0.007$ & $3.8\pm1.9$ & $\sim0.02$ & 30\%  \\
$f_\mathrm{mol}>$ 0.16 & $130\pm3$ & $1.3\pm0.05$ & $30\pm20$ & $\sim0.08$ & 40\%  \\
    \hline\\
    \end{tabular}
    \caption*{\textbf{Notes.} Physical properties of the outflows in the sample derived from the high-resolution spectra of the \ha line.
    The first part of the table shows outflow velocities, extents, mass rates and mass loading factors ($\eta=\dot{M}_\mathrm{out}/\mathrm{SFR}$) for individual galaxies in the sample with outflows.
    Parts two to six show the mean values of the complete sample with outflow detections, and split into bins according to their SFR, $M_*$, and molecular gas fraction $f_\mathrm{mol}$.
    The demarcation value between the two bins for each quantity was chosen as the median value of that quantity  in the outflow sample.
    For the bins we also show the corresponding outflow incidence fraction in the last column.
    We mark results as approximate ($\sim$) where the statistical uncertainty matches or exceeds the derived value.}
    \label{tab:outflow_props}
\end{table*}

\subsection{Observed outflow incidence in the local universe}
\label{sec:dis:higmass_obs_comp}
Previous studies of the local galaxy population generally report outflows to be common in star-forming main-sequence galaxies with $M_*\gtrsim10^{10}\,\mathrm{M_\odot}$ \citep[e.g.][]{chenABSORPTIONLINEPROBESPREVALENCE2010, ciconeOutflowsComplexStellar2016, averyIncidenceScalingRelations2021}.
\citet{chenABSORPTIONLINEPROBESPREVALENCE2010} use nebular sodium doublet (NaI D) absorption to identify outflows and find incidence fractions of around 30\% at the high-mass end of their sample.
Similarly, \citet{roberts-borsaniPrevalencePropertiesCold2019}, report a prevalence of neutral atomic outflows in star-forming galaxies with $M_*\gtrsim10^{10}\,\msun$ based on their stacking analysis of nebular sodium absorption in SDSS DR7 galaxies.
Based on MaNGA integral field spectroscopy, \citet{averyIncidenceScalingRelations2021} identify outflows in 12\% of their sample, with the vast majority of outflows identified in galaxies with $M_*\gtrsim10^{10}\,\mathrm{M_\odot}$, where the incidence fraction is $>20\%$.
Both of these studies report a sharp drop in the outflow incidence below $M_*\sim10^{10}\,\mathrm{M_\odot}$, with incidence fractions of only a few percent at lower masses.
Similarly, \citet{concasLightBreezeLocal2017} find no evidence of outflows in galaxies with $M_*<10^{10}\,\mathrm{M_\odot}$ in their analysis of the [OIII] line in stacked SDSS spectra.

Our results show that the outflow incidence in our sample of low-mass galaxies is strikingly similar to that at higher masses, when measured based on high-resolution spectroscopy, indicating that the drop-off observed in previous studies based on optical emission lines may be largely due to insufficient spectral resolution to resolve outflows in the lower mass regime.
Indeed, quite strikingly, we have demonstrated that already at $M_*\sim10^{10}\,\mathrm{M_\odot}$ the SDSS spectroscopy does not fully resolve optical emission line profiles, indicating that any outflow studies (or line profile studies) in this mass regime are hampered by resolution effects.
For studies using sodium absorption to trace the neutral atomic phase, the difference in outflow incidence at low and high masses could be due to other effects, such as the absence of dust making NaI D ineffective as a tracer \citep{roberts-borsaniPrevalencePropertiesCold2019}.

Outflows in low-mass galaxies have been successfully detected with high-resolution spectroscopy in small samples of local galaxies \citep{manzano-kingAGNDrivenOutflowsDwarf2019, xuEMPRESSVIOutflows2022, marascoShakenNotExpelled2023}.
The low-velocity nature of the outflows described in \citet{xuEMPRESSVIOutflows2022} and \citet{marascoShakenNotExpelled2023} also supports the explanation that their kinematic signatures would not be detectable with lower resolution spectroscopy, leading to the trend with stellar mass observed in other studies \citep{chenABSORPTIONLINEPROBESPREVALENCE2010, concasLightBreezeLocal2017, averyIncidenceScalingRelations2021}.
\citet{manzano-kingAGNDrivenOutflowsDwarf2019} find signatures of outflows in 13/50 ($26\%$) galaxies in their mixed sample of objects with and without indicators of AGN ionization and masses between $10^8\mathrm{M_\odot}$ and $10^{10}\,\mathrm{M_\odot}$ observed with Keck/LRIS with a spectral resolution of $\sigma\sim80$\,km/s.
This is similar to the incidence we find here, although most of the outflow candidates in their sample are associated with AGN activity, whereas none of the outflow candidates in our sample show clear signs of AGN activity.

\subsection{Comparison to theoretical predictions for low-mass galaxies}
\label{sec:dis:lowmass_theory_comp}
In this section we discuss the outflow incidence in our sample in light of the the ubiquity of galactic winds invoked to explain a number of phenomena such as the observed stellar mass to halo mass relation \citep{behrooziCOMPREHENSIVEANALYSISUNCERTAINTIES2010}, low metal fraction of dwarf galaxies \citep{mcquinnLEOHOWMANY2015}, and properties of their dark matter halos \citep{navarroCoresDwarfGalaxy1996, governatoBulgelessDwarfGalaxies2010, governatoCuspyNoMore2012, teyssierCuspcoreTransformationsDwarf2013}.
An incidence fraction of $\sim30\%$ may seem low in light of these predictions, but it is important to keep in mind the limitations of observational techniques for the identification of outflows.
Spectroscopic identification of outflows depends on the presence of gas with line-of-sight velocities distinct from those of quiescent gas in the ISM.

If a typical outflow has a bi-conical morphology with an opening angle of $\theta=45^\circ$, which is consistent with detailed observations in the local universe \citep{venturiMAGNUMSurveyMUSE2018, lopez-cobaAMUSINGNearbyGalaxy2020}, it subtends a solid angle of $\Omega_\mathrm{out} = 4\pi(1-\cos\theta)\approx3.7$.
The probability of observing outflowing gas in the line-of-sight can then be approximated as $\Omega_\mathrm{out}/(4\pi)\approx0.3$.
An observed outflow incidence fraction of $\sim30\%$ would, with these assumptions, be consistent with outflows being near-ubiquitous in the sample.
This interpretation is in line with the argument made by \citet{carnianiJADESIncidenceRate2023} who applied the same considerations to a sample of low-mass galaxies at higher redshift and reported an outflow incidence of 25\%-40\%.

The timescales of star formation and outflow lifetimes may also have an effect on the observed incidence.
Even if all galaxies are affected by outflows over the course of their evolution, these events could be relatively short-lived in galaxies where star formation happens in short bursts.

\subsection{Mass loading}
\label{sec:dis:mass_loading}
In this section we estimate the mass loading factor $\eta=\dot{M}_\mathrm{out}/\mathrm{SFR}$ of outflows observed in our sample.
As we have no information on the extent of the outflows, we need to assume one.
The results should therefore be treated as order-of-magnitude estimates.

Since our higher-resolution spectra are not calibrated in terms of absolute flux, we combine them with integrated line fluxes derived from SDSS spectra to estimate outflow masses.
To this end, we multiply the fraction of the flux contained in the outflow component(s) (see Section~\ref{sec:res:outflows} and Table~\ref{tab:highres_info}) by the integrated line flux of the \ha line measured in the SDSS fiber.
We adopt \ha fluxes from \citet{hagedornMolecularGasScaling2024}, where they are derived from a full-spectrum fit to the SDSS spectra using the \texttt{GandALF} code \citep{sarziSAURONProjectIntegralfield2006}, which takes into account \ha absorption in the stellar continuum and dust extinction.
After converting the resulting integrated outflow fluxes to luminosities, we obtain outflow masses following \citet{fluetschPropertiesMultiphaseOutflows2021}.

As an estimate for the outflow velocity, we use $|v02|$ (see Section \ref{sec:res:asymmetry_wing_tests}).
Finally, we estimate the physical extent of the outflow to be comparable to the size of the galaxy, adopting $r_{50}$, the radius containing 50\% of the Petrosian r-band flux, obtained from SDSS DR8 \citep{AiharaSDSSdr8}, as the outflow extent.
We then calculate outflow mass rates as:
$$\dot{M}_\mathrm{out} = \frac{M_\mathrm{out}\times|v02|}{r_{50}}.$$
This results in a median outflow mass rate of $\dot{M}_\mathrm{out}=10^{-2}\,\mathrm{M_\odot/yr}$ and a mean mass loading factor of $\sim0.07$.
Table~\ref{tab:outflow_props} lists the outflow properties for individual galaxies in the sample.

This is about an order of magnitude lower than what is reported in \citet{mcquinnGalacticWindsLowmass2019}, but consistent with observations by \citet{marascoShakenNotExpelled2023}, both of which use \ha to trace the warm ionized outflows.
The latter point out that these discrepancies are within what is expected from differences in methods used to measure outflows.

Comparing the mass loading factors derived here with theoretical predictions is not trivial.
Our observation measure only the warm ionized phase of the gas, whereas cosmological simulations generally do not distinguish different gas phases and are therefore likely to report higher mas outflow rates than observations of a single tracer.
Based on the EAGLE simulations, \citet{mitchellGalacticOutflowRates2020} predict mass loading factors of $\sim1-3$ in the mass range of our sample, exceeding our results by at least an order of magnitude.
Similarly, \citet{nelsonFirstResultsTNG502019} predict $\eta\sim3-10$, based on the TNG50 simulations, about 2 orders of magnitude higher than what we find here.
This either implies that the warm ionized phase of the outflow accounts for only about 1\% of the mass outflow rate in these galaxies, that theoretical feedback recipes lead to more extreme outflows than observed, or a combination thereof.

Observations of outflows in ultra-luminous infrared galaxies (ULIRGs) using multiple tracers show mass rate contributions $<1\%$ in these objects \citep{fluetschPropertiesMultiphaseOutflows2021}.
The well-studied outflow in the local starburst M82 shows a similar partition, with the warm ionized phase contributing only a small fraction of the mass outflow rate \citep{shopbellAsymmetricWindM821998} compared to the neutral atomic \citep{contursiSpectroscopicFIRMapping2013} and cold molecular \citep{leroyMULTIPHASECOLDFOUNTAIN2015} phases, consistent with the $\sim1\%$ our results imply.
\citet{romanoStarformationdrivenOutflowsLocal2023} report mass loading factors of the order of unity in the neutral atomic phase of outflows in a sample of low-mass galaxies, which would be consistent with a small contribution from the ionized phase in combination with our estimates.
Due to the generally low S/N of the available CO observations, we are not able to investigate the presence of cold gas counterparts to the ionized outflows in our sample directly.

We test the dependence of outflow properties in the sample on SFR, $M_*$, and molecular gas fraction $f_\mathrm{mol}=M_\mathrm{mol}/M_*$ by splitting the sample into two bins for each of these quantities.
To demarcate the bins we use the median of the outflowing sample in each case (SFR, $M_*$, and $f_\mathrm{mol}$).
We adopt molecular gas fractions for the 8 galaxies with CO detections from \citet{hagedornMolecularGasScaling2024}.
The results are listed in Table~\ref{tab:outflow_props}.
Outflow velocities are consistent across the bins, given the associated statistical uncertainties.
Uncertainties on the derived outflow properties, mass rate and mass loading, are generally high, so that no significant differences can be determined between bins.
The most pronounced difference is in the mass outflow rate of galaxies with $f_\mathrm{mol}>0.16$ vs those with  $f_\mathrm{mol}<0.16$, the former showing a higher mass outflow rate, if only at $\sim1\sigma$ significance.

Table~\ref{tab:outflow_props} also shows outflow incidence fractions (compared to all galaxies with \ha S/N$>$10 falling into the relevant bin) for the different bins described above.
Incidence is enhanced in the high-SFR, whereas differences are negligible in the $M_*$ and $f_\mathrm{mol}$ bins, given the size of the sample.
The behavior with SFR is not surprising for a regime in which we expect primarily star-formation driven outflows.
Following the arguments made in Section~\ref{sec:dis:lowmass_theory_comp}, the enhanced incidence for high-SFR galaxies may also imply larger average outflow opening angles in this regime.

\section{Summary and conclusion}
\label{sec:conclusion}
We present high-resolution TNG and NTT spectroscopy for 52 local galaxies with stellar masses below $10^{10}\,\mathrm{M_\odot}$.
Comparing gas velocity dispersions inferred from the \ha and [OIII] line profiles in the high-resolution data to those inferred from SDSS data shows that the SDSS spectroscopy does not resolve emission line profiles in these objects.
As a result, asymmetries and broad wings evident in the high-resolution spectra, quantified by excess kurtosis and non-parametric velocity measures, are largely absent in the low-resolution SDSS spectroscopy, where line profiles are dominated by the instrumental dispersion profile.

We investigate whether these features correspond to potential galactic outflows and find that $\sim30\%$ of galaxies in our sample with \ha or [OIII] emission lines detected at S/N$>$10 show signatures of potential outflows, which account for between 10\% and 40\% of the total integrated line flux.
The lack of outflow detections at stellar masses below $10^{10}\,\mathrm{M_\odot}$  in local galaxies reported in some studies based on optical emission lines \citep[e.g.][]{ciconeOutflowsComplexStellar2016, concasLightBreezeLocal2017, averyIncidenceScalingRelations2021} is therefore likely due to the use of low-spectral resolution data, in which the narrow emission lines of low-mass galaxies are not sufficiently resolved to identify outflow signatures.
Based on the fraction of the line flux attributed to the outflows, we estimate the median mass loading factor in the outflows sample to be $\sim0.07$.
This is consistent with predictions from numerical simulations \citep{nelsonFirstResultsTNG502019, mitchellGalacticOutflowRates2020} if the warm ionized phase of the outflow contributes about 1\% of the total mass loading, which is supported by multi-phase observations \citep{fluetschPropertiesMultiphaseOutflows2021, heckmanGalacticWindsRole2017}.

Assuming that outflows in the low-mass regime present predominantly as bi-conical structures with opening angles of $45^{\circ}$, the $\sim30\%$ outflow incidence we find in our sample suggests outflows are near-ubiquitous in the low-mass regime.
This is consistent with predictions from theoretical works, which invoke such outflows to regulate star-formation \citep{dekelOriginDwarfGalaxies1986, lowStarburstdrivenMassLoss1999, ferraraRoleStellarFeedback2000, behrooziCOMPREHENSIVEANALYSISUNCERTAINTIES2010, daveGalaxyEvolutionCosmological2011}, eject metals from the ISM \citep{mcquinnLEOHOWMANY2015, brooksOriginEvolutionMassMetallicity2007}, and shape the dark matter halos \citep{navarroCoresDwarfGalaxy1996, governatoBulgelessDwarfGalaxies2010, governatoCuspyNoMore2012, teyssierCuspcoreTransformationsDwarf2013} in low-mass galaxies.
We observe enhanced incidence fractions for galaxies with above average SFR and \ha S/N in the sample, as expected for star-formation driven outflows.

\begin{acknowledgements}
    We thank the anonymous referee for their insightful comments.
    This work makes use of data from the Sloan Digital Sky Survey (SDSS).
    Funding for SDSS-III has been provided by the Alfred P. Sloan Foundation, the Participating Institutions, the National Science Foundation, and the U.S. Department of Energy Office of Science. The SDSS-III web site is http://www.sdss3.org/.
    SDSS-III is managed by the Astrophysical Research Consortium for the Participating Institutions of the SDSS-III Collaboration including the University of Arizona, the Brazilian Participation Group, Brookhaven National Laboratory, Carnegie Mellon University, University of Florida, the French Participation Group, the German Participation Group, Harvard University, the Instituto de Astrofisica de Canarias, the Michigan State/Notre Dame/JINA Participation Group, Johns Hopkins University, Lawrence Berkeley National Laboratory, Max Planck Institute for Astrophysics, Max Planck Institute for Extraterrestrial Physics, New Mexico State University, New York University, Ohio State University, Pennsylvania State University, University of Portsmouth, Princeton University, the Spanish Participation Group, University of Tokyo, University of Utah, Vanderbilt University, University of Virginia, University of Washington, and Yale University. 
    CC acknowledges funding from the European Union's Horizon Europe research and innovation programme under grant agreement No. 101188037 (AtLAST2).
    CC, CV, PS acknowledge financial support from the INAF Bando Ricerca Fondamentale INAF 2022 Large Grant: “Dual and binary supermassive black holes in the multi-messenger era: from galaxy mergers to gravitational waves” and from the INAF Bando Ricerca Fondamentale INAF 2024 Large Grant: “The Quest for dual and binary massive black holes in the gravitational wave era”.
    %
\end{acknowledgements}

\bibliography{Project2_sources}


\appendix

\section{SDSS vs NTT/TNG line widths}
\label{app:linewidths}
\begin{figure}
    \centering
    \includegraphics[clip=true, trim=0.4cm 0.3cm 0.3cm 0.3cm, width=0.48\textwidth]{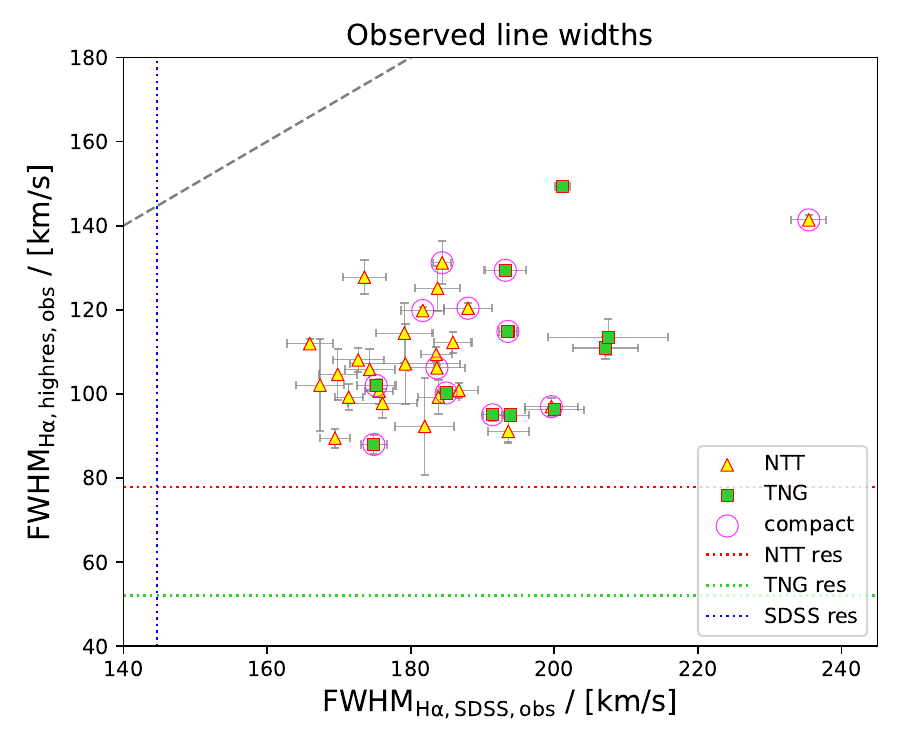}\quad
    \includegraphics[clip=true, trim=0.4cm 0.3cm 0.3cm 0.3cm, width=0.48\textwidth]{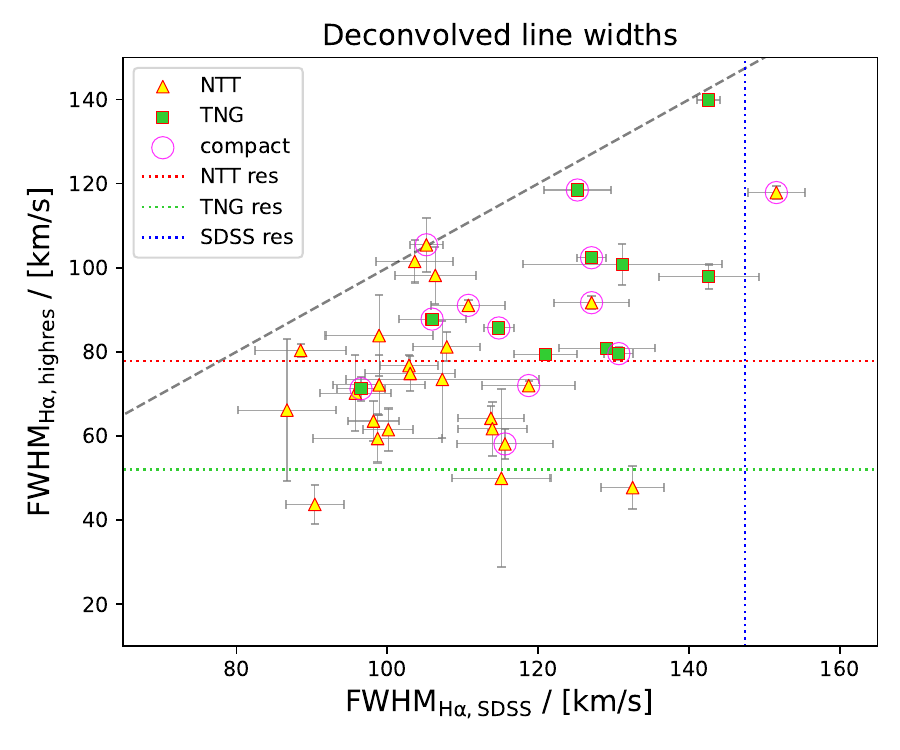}\\
    \caption{Comparison between H$\alpha$ line widths measured using SDSS and TNG/NTT. The top panel shows the observed values, while the bottom panel shows the values obtained after subtracting in quadrature the instrumental resolution. The average resolution for the three instruments is shown by colored dotted lines for reference. Magenta circles mark 12 particularly compact galaxies, for which over 30 percent of r-band flux from a 10'' aperture is contained within the 3'' SDSS fiber. The gray dashed line shows the one-to-one relation. We show only galaxies in which \ha is detected with an S/N greater than 3 and for which the uncertainty on the best-fit line width does not exceed the line widths itself. We chose these thresholds to exclude galaxies in which the line width cannot be determined accurately.}
    \label{fig:fwhm_comp}
\end{figure}

Here, we compare the line widths of \ha obtained from a single-component Gaussian template fit, as described in Section~\ref{sec:methods:line_fitting}, to the SDSS and higher-resolution spectra.
Fig.~\ref{fig:fwhm_comp} shows \ha line widths based SDSS data vs those inferred from high-resolution spectroscopy for the sample.
When not accounting for instrumental resolution, the line widths inferred from the high-resolution spectroscopy are noticeably lower than for the same lines observed with SDSS, as expected from the differences in instrumental resolution.
After deconvolving the line profiles from the instrumental resolution measured for each dataset, this difference shrinks, as expected, but does not disappear completely.
The SDSS inferred line widths remain systematically higher than those inferred from high-resolution spectroscopy.
This could be due to an inaccurate estimate of the instrumental resolution based on arc-lamp spectra, or due to the different apertures between the SDSS and high-resolution observations.
To investigate the second possibility, we select galaxies in our sample that are particularly compact based on their SDSS r-band imaging.
We consider a galaxy compact relative to the 3'' SDSS fiber size, if the fiber aperture contains more than 30\% of the r-band flux contained in a synthetic 10'' aperture.
The 12 galaxies fulfilling this criterion are marked in Fig.~\ref{fig:fwhm_comp} by magenta circles, and they show the same trends as the parent sample, leading us to conclude that aperture effects are not dominant in this case.
This suggests that the deconvolution method does not produce accurate line widths, even for marginally resolved SDSS spectra.

\onecolumn
\section{NTT/EFOSC2 line spectra}
\label{app:ntt_specs}

\begin{figure}[h]
    \centering
\includegraphics[clip=true,trim=0.8cm 0 0.8cm 0.3,width=0.32\textwidth]{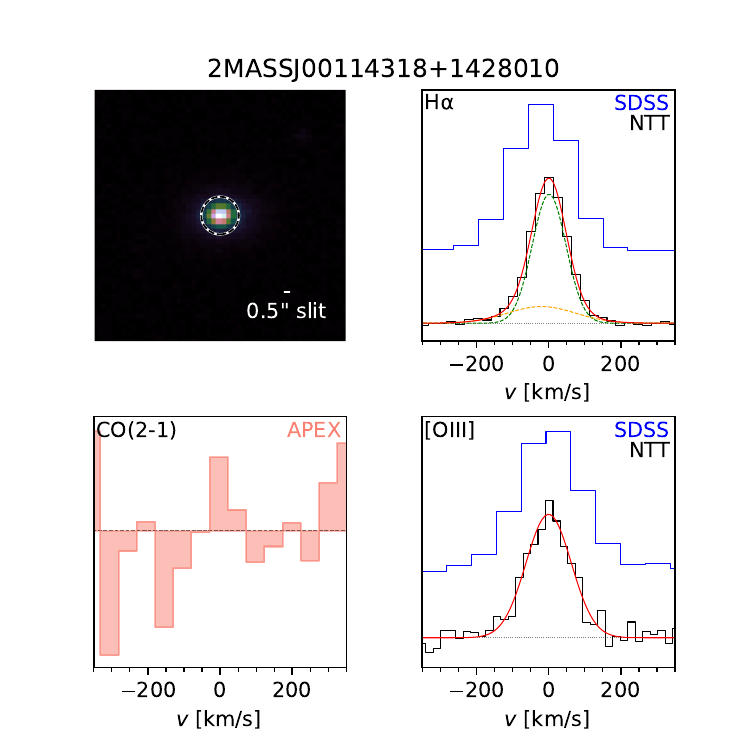}
\includegraphics[clip=true,trim=0.8cm 0 0.8cm 0.3,width=0.32\textwidth]{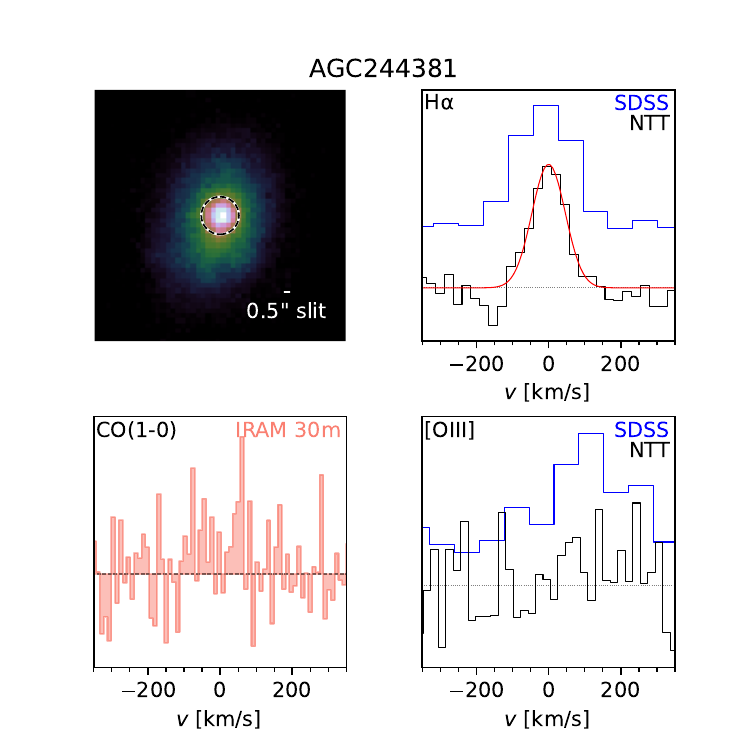}
\includegraphics[clip=true,trim=0.8cm 0 0.8cm 0.3,width=0.32\textwidth]{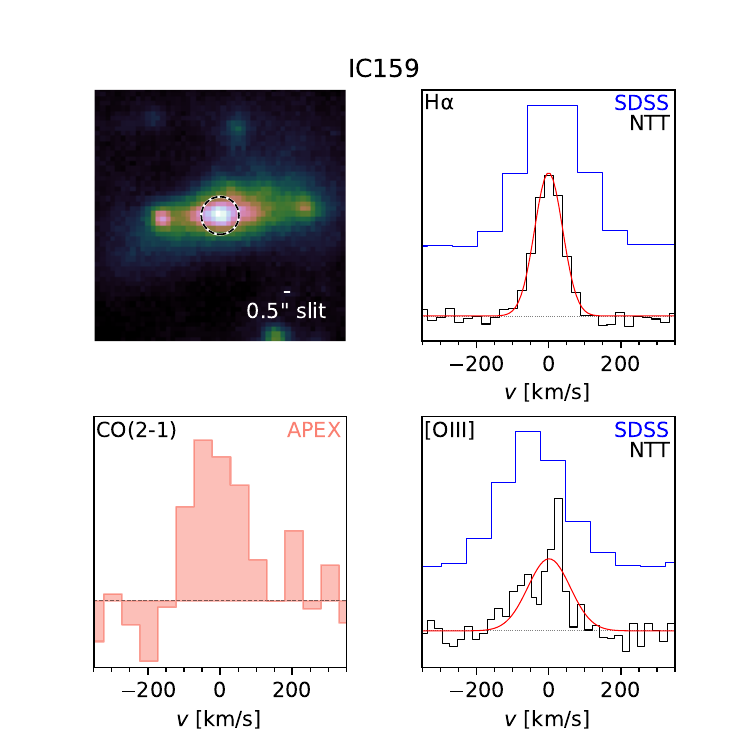}
\includegraphics[clip=true,trim=0.8cm 0 0.8cm 0.3,width=0.32\textwidth]{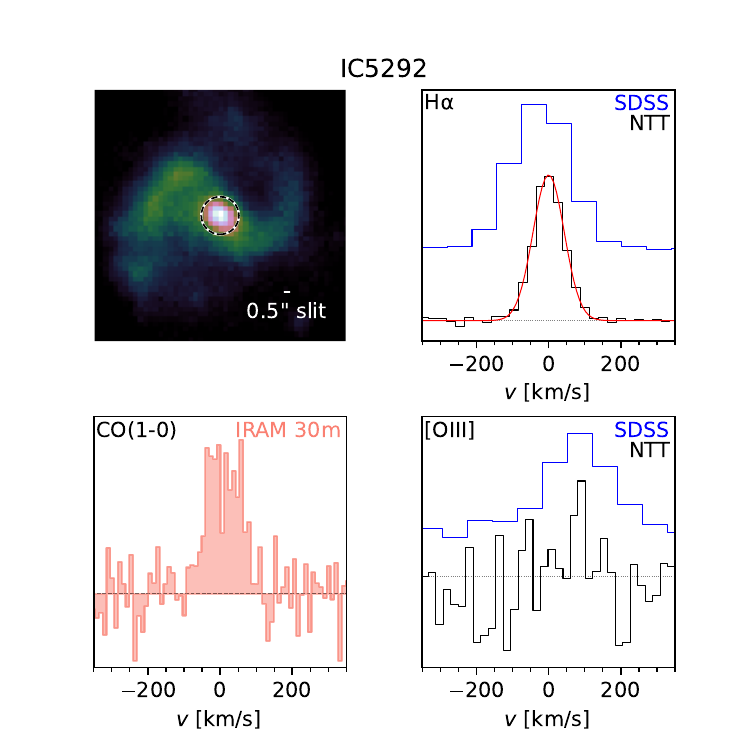}
\includegraphics[clip=true,trim=0.8cm 0 0.8cm 0.3,width=0.32\textwidth]{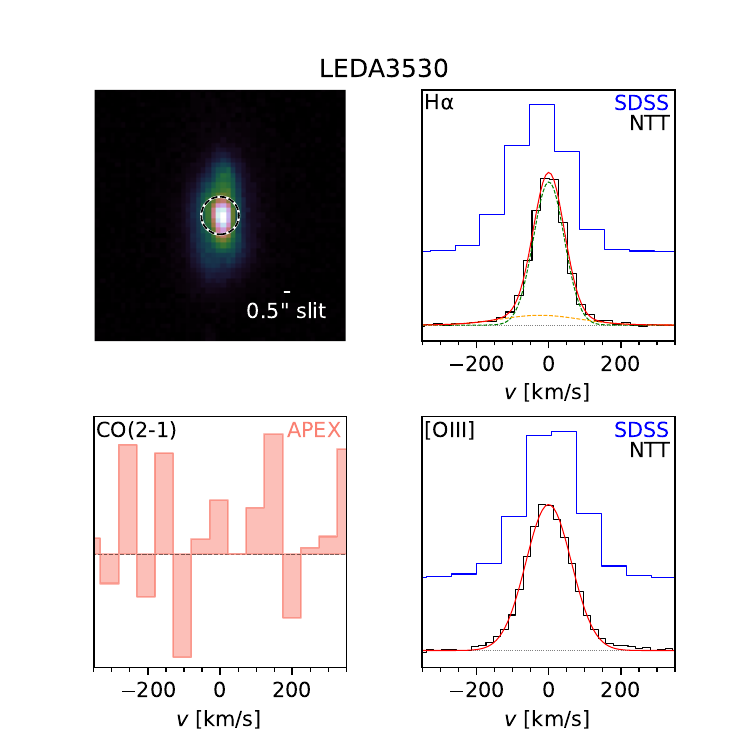}
\includegraphics[clip=true,trim=0.8cm 0 0.8cm 0.3,width=0.32\textwidth]{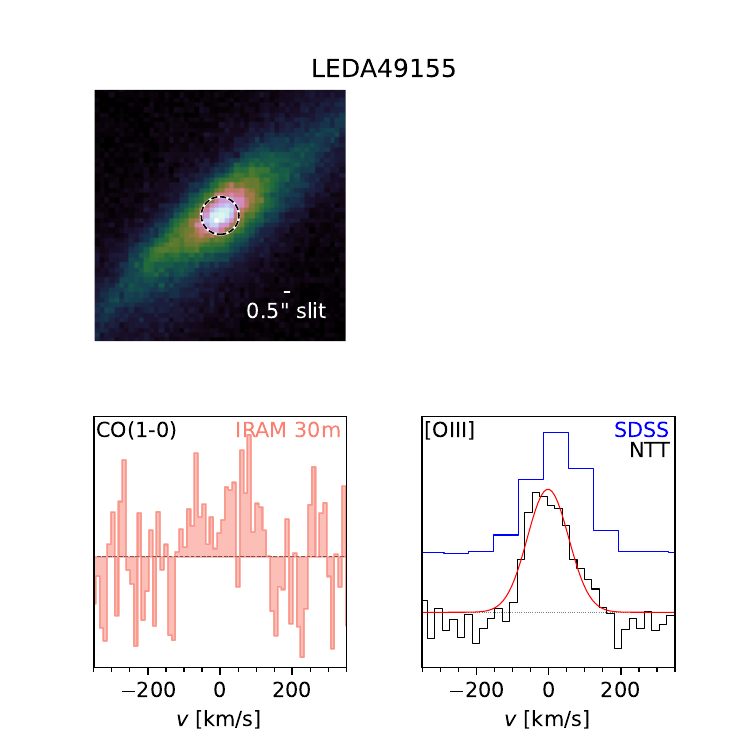}
    \caption{Targets observed with NTT/EFOSC2.  
    The image on the top left in each panel shows the SDSS r-band image of the target, with the size and position of the 3'' SDSS fiber shown as a black-and-white dashed circle. 
    The white bar in each image shows the size of the slit used in the NTT/EFOSC2 spectroscopy.
    On the top right in each panel we show the SDSS (on top in blue) and higher resolution NTT/EFOSC2 (below in black) spectra around the H$\alpha$ emission line at 6562.8\,\AA. 
    On the bottom right in each panel is the same plot for the [OIII] emission line at 5006.77\,\AA.
    For emission lines detected with S/N$>$3 we also show the best fit model in red.
    If the model has multiple components, individual components are shown as dashed curves in green (non-outflow components) and orange (outflow components).
    On the bottom left is the region around the CO(1-0) or CO(2-1) emission line, whichever was used in the analysis, from the ALLSMOG and xCOLDGASS surveys.
    This figure is continued below in Figs.~\ref{fig:ha_cards_ntt_2}, \ref{fig:ha_cards_ntt_3}, \ref{fig:ha_cards_ntt_4}, and \ref{fig:ha_cards_ntt_5}.}
    \label{fig:ha_cards_ntt_1}
\end{figure}

\onecolumn
\begin{figure}
    \centering
\includegraphics[clip=true,trim=0.8cm 0 0.8cm 0.3,width=0.32\textwidth]{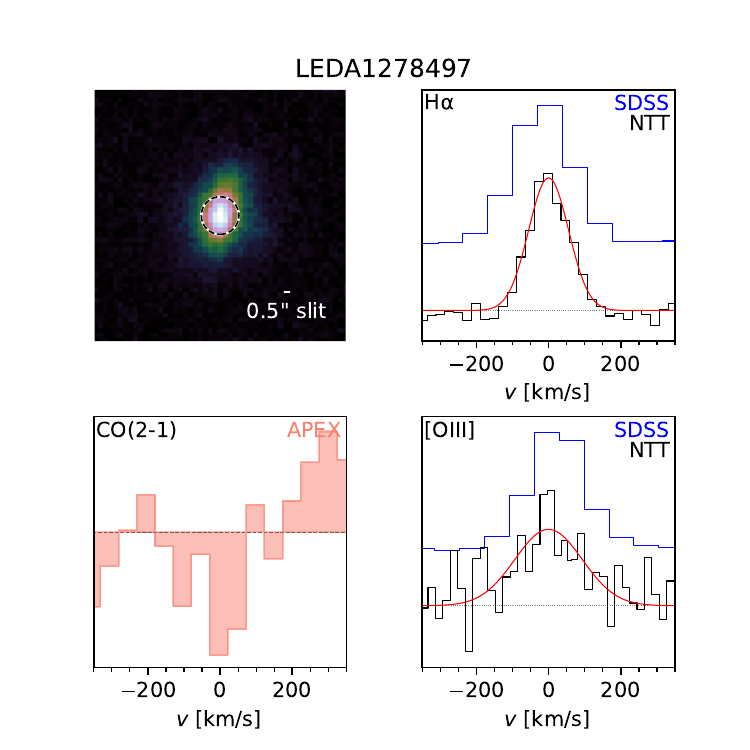}
\includegraphics[clip=true,trim=0.8cm 0 0.8cm 0.3,width=0.32\textwidth]{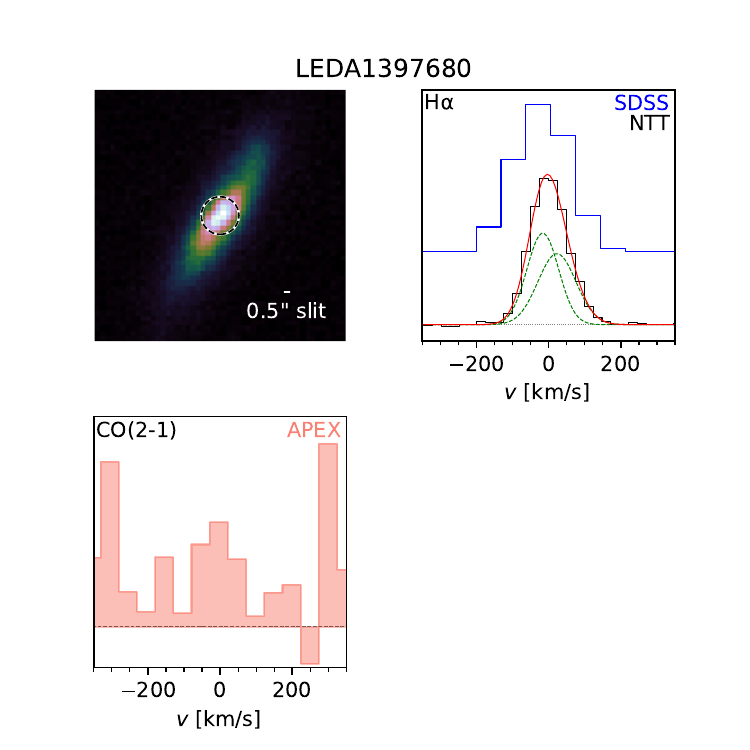}
\includegraphics[clip=true,trim=0.8cm 0 0.8cm 0.3,width=0.32\textwidth]{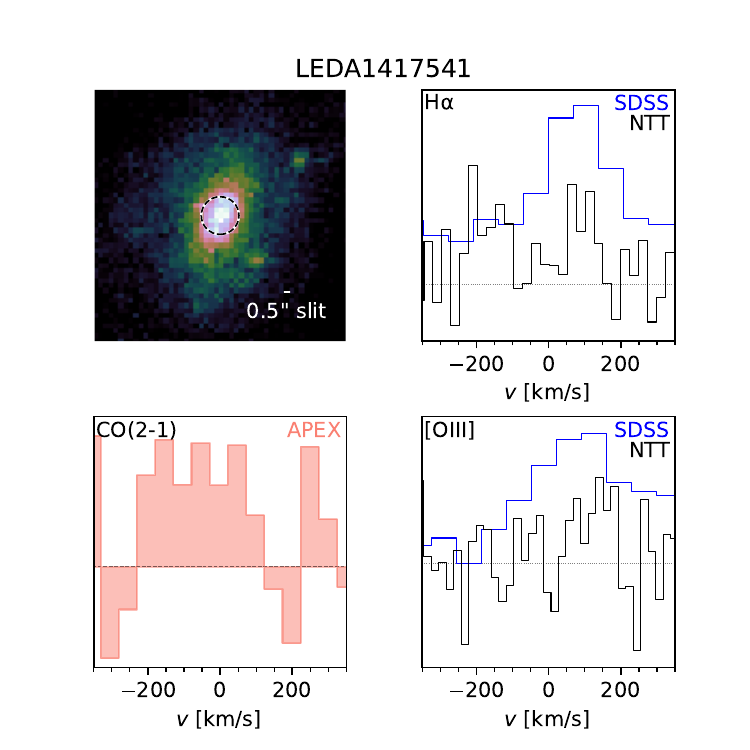}
\includegraphics[clip=true,trim=0.8cm 0 0.8cm 0.3,width=0.32\textwidth]{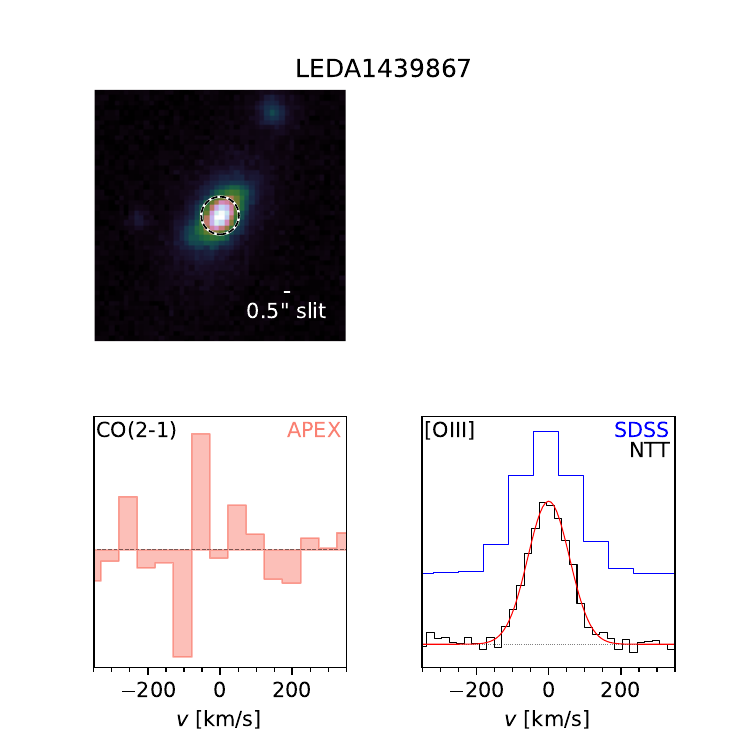}
\includegraphics[clip=true,trim=0.8cm 0 0.8cm 0.3,width=0.32\textwidth]{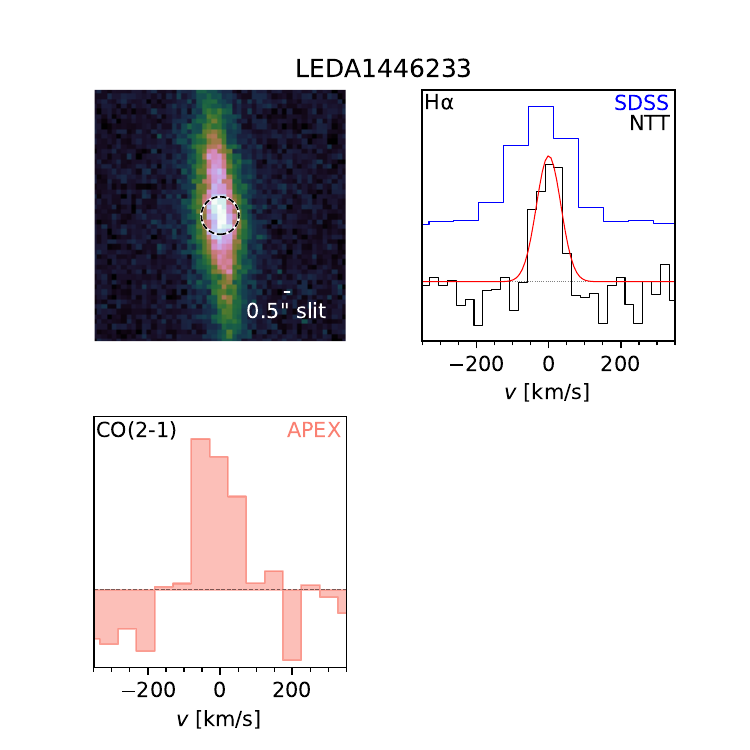}
\includegraphics[clip=true,trim=0.8cm 0 0.8cm 0.3,width=0.32\textwidth]{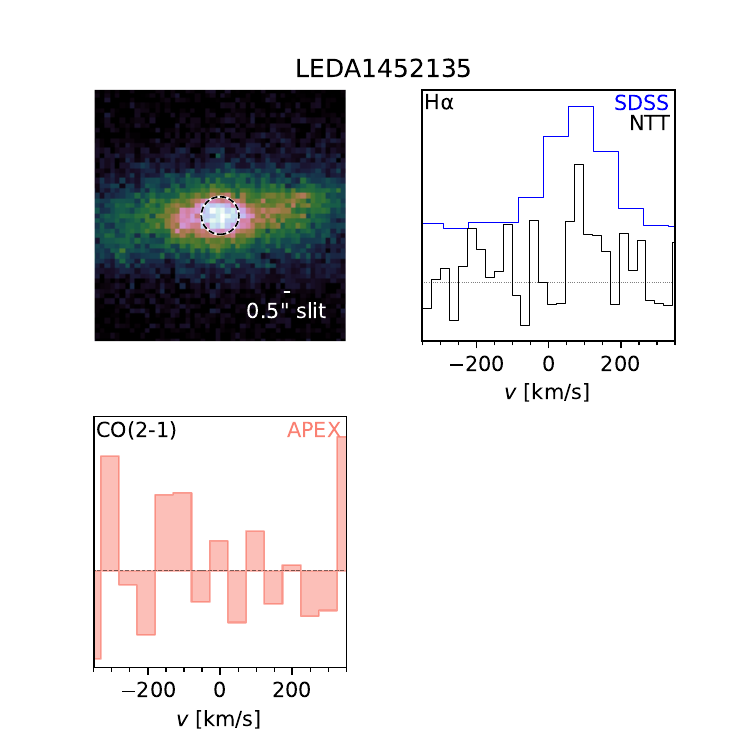}
\includegraphics[clip=true,trim=0.8cm 0 0.8cm 0.3,width=0.32\textwidth]{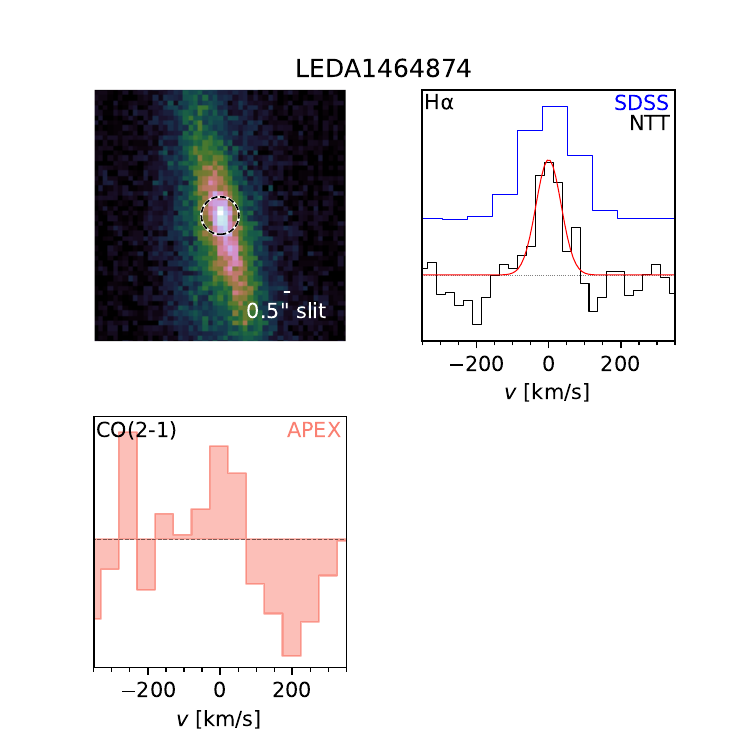}
\includegraphics[clip=true,trim=0.8cm 0 0.8cm 0.3,width=0.32\textwidth]{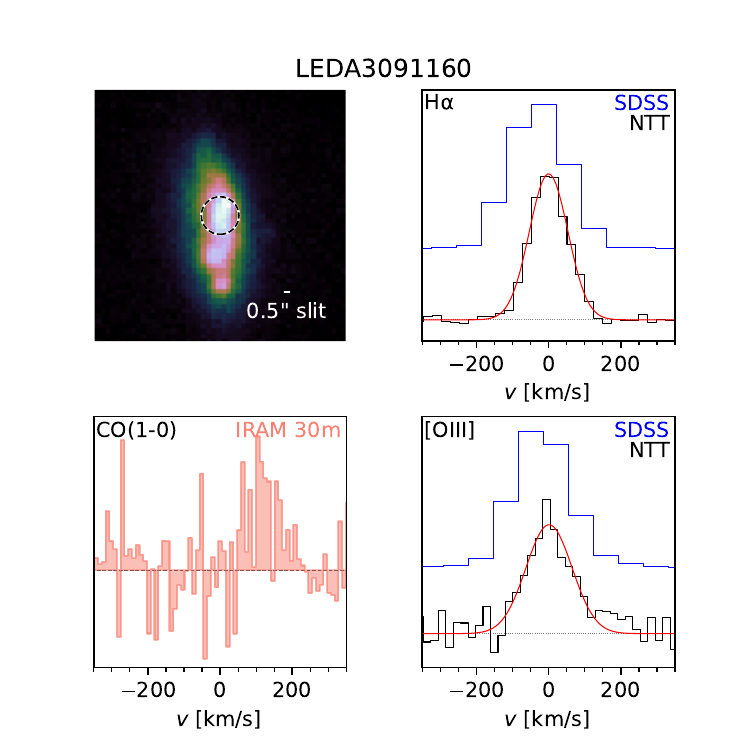}
\includegraphics[clip=true,trim=0.8cm 0 0.8cm 0.3,width=0.32\textwidth]{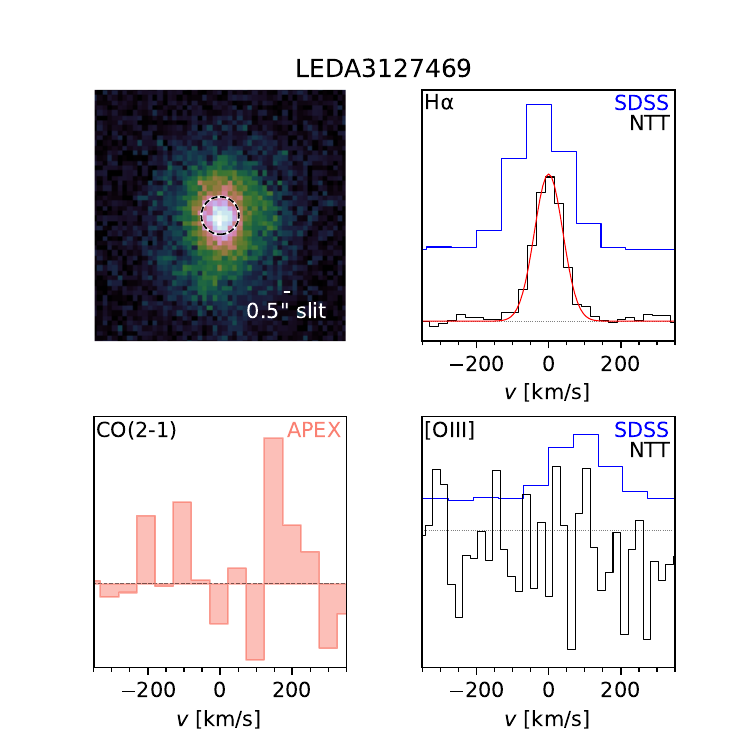}
    \caption{Fig.~\ref{fig:ha_cards_ntt_1} cont.}
    \label{fig:ha_cards_ntt_2}
\end{figure}

\onecolumn
\begin{figure}
    \centering
\includegraphics[clip=true,trim=0.8cm 0 0.8cm 0.3,width=0.32\textwidth]{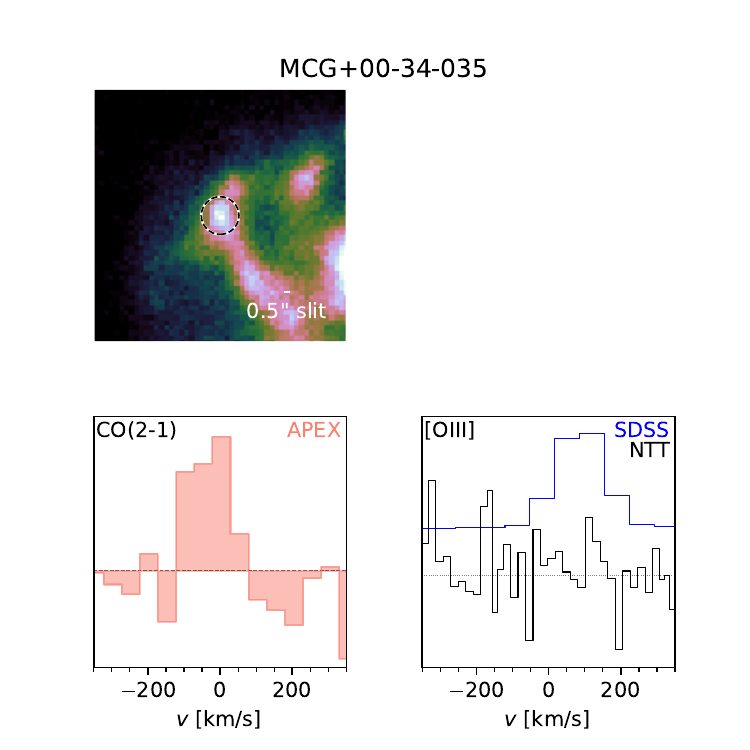}
\includegraphics[clip=true,trim=0.8cm 0 0.8cm 0.3,width=0.32\textwidth]{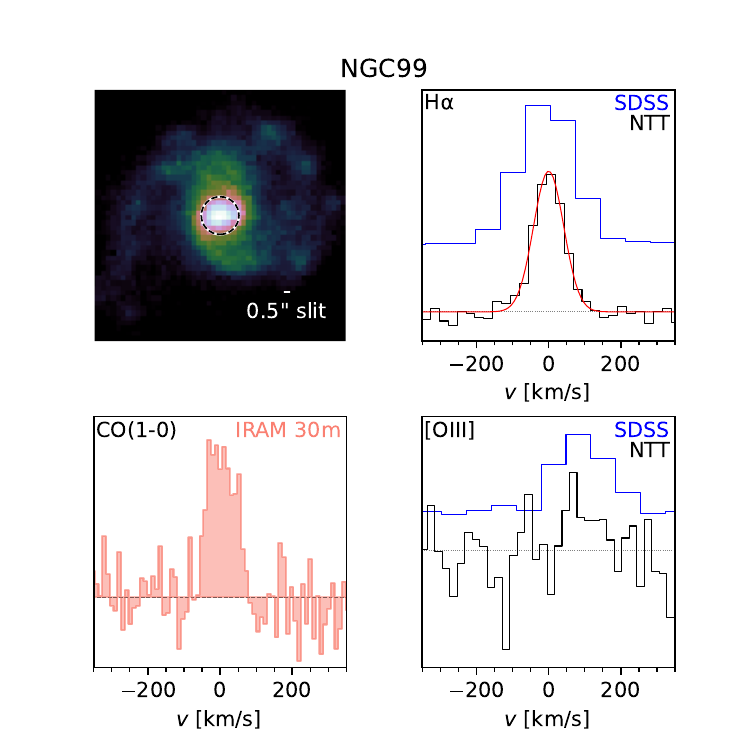}
\includegraphics[clip=true,trim=0.8cm 0 0.8cm 0.3,width=0.32\textwidth]{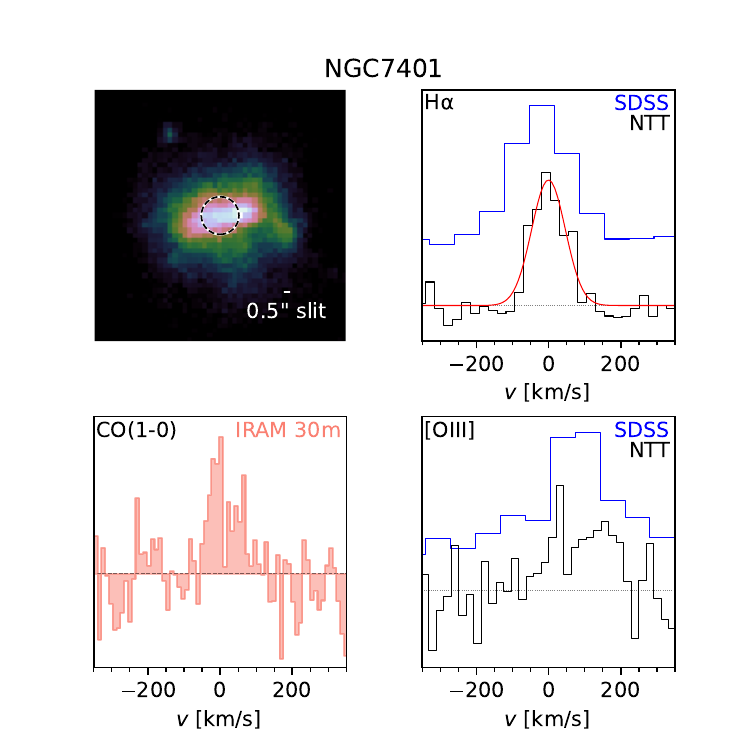}
\includegraphics[clip=true,trim=0.8cm 0 0.8cm 0.3,width=0.32\textwidth]{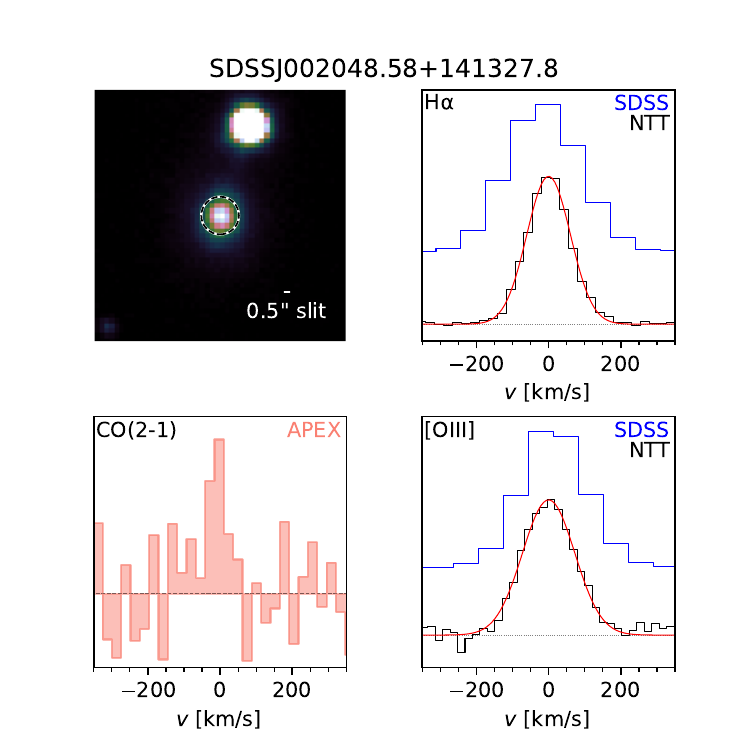}
\includegraphics[clip=true,trim=0.8cm 0 0.8cm 0.3,width=0.32\textwidth]{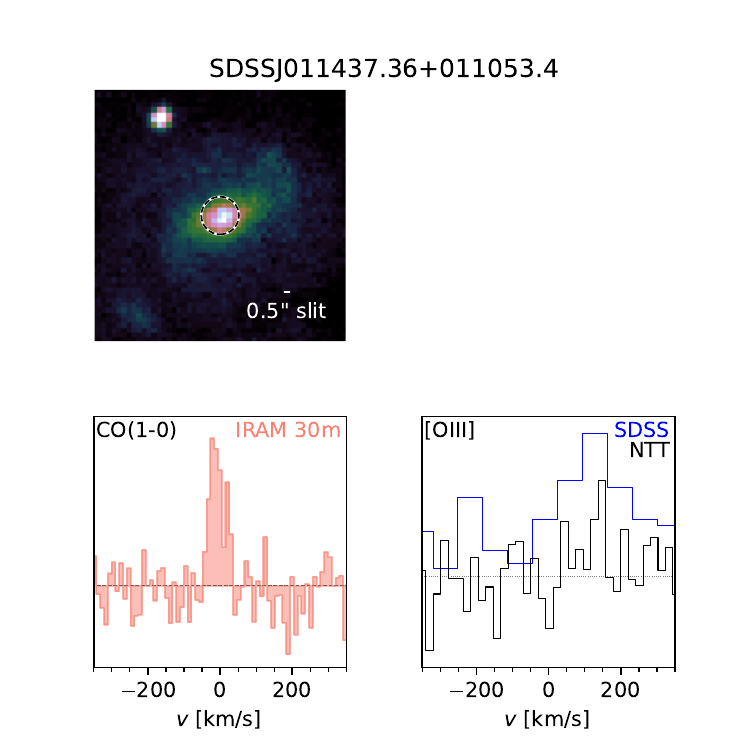}
\includegraphics[clip=true,trim=0.8cm 0 0.8cm 0.3,width=0.32\textwidth]{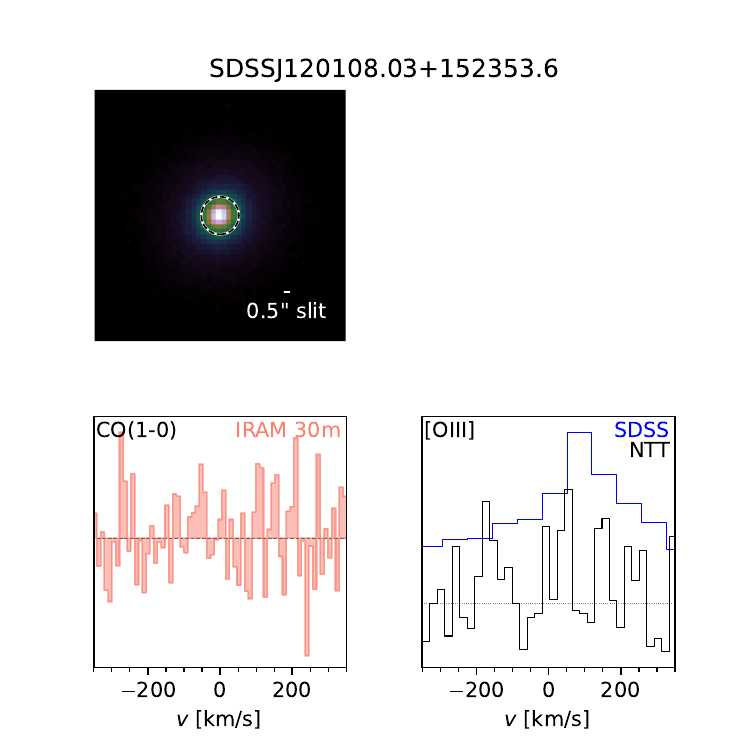}
\includegraphics[clip=true,trim=0.8cm 0 0.8cm 0.3,width=0.32\textwidth]{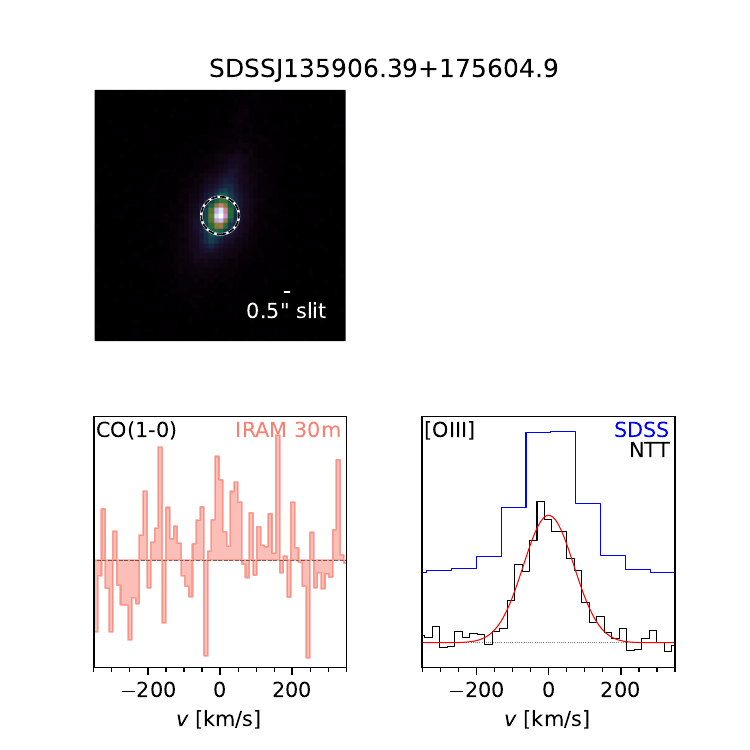}
\includegraphics[clip=true,trim=0.8cm 0 0.8cm 0.3,width=0.32\textwidth]{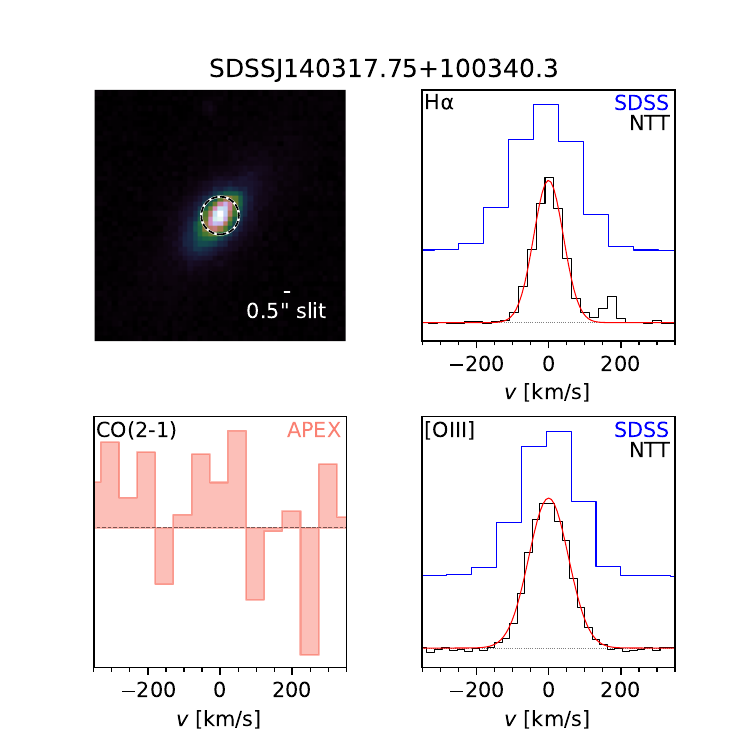}
\includegraphics[clip=true,trim=0.8cm 0 0.8cm 0.3,width=0.32\textwidth]{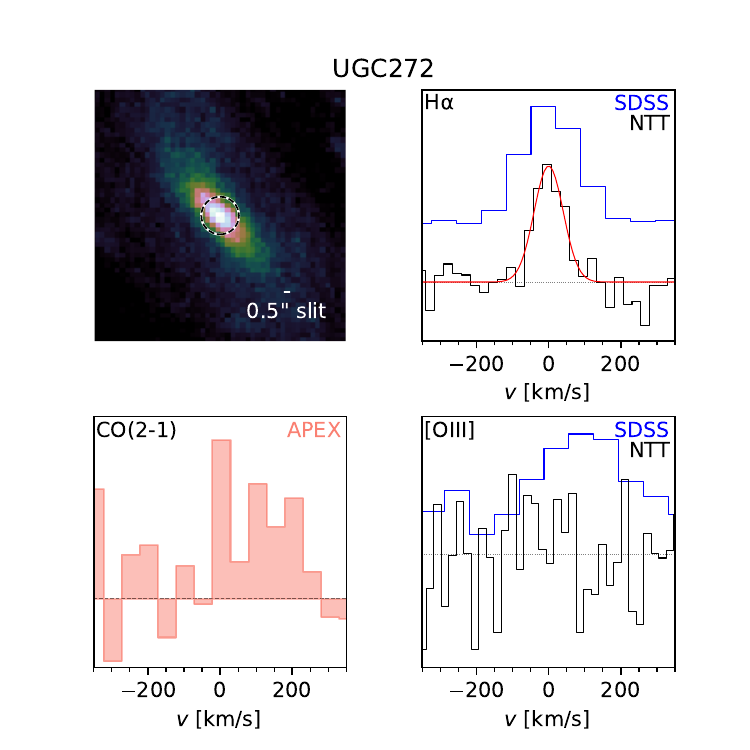}
    \caption{Fig.~\ref{fig:ha_cards_ntt_1} cont.}
    \label{fig:ha_cards_ntt_3}
\end{figure}

\onecolumn
\begin{figure}
    \centering
\includegraphics[clip=true,trim=0.8cm 0 0.8cm 0.3,width=0.32\textwidth]{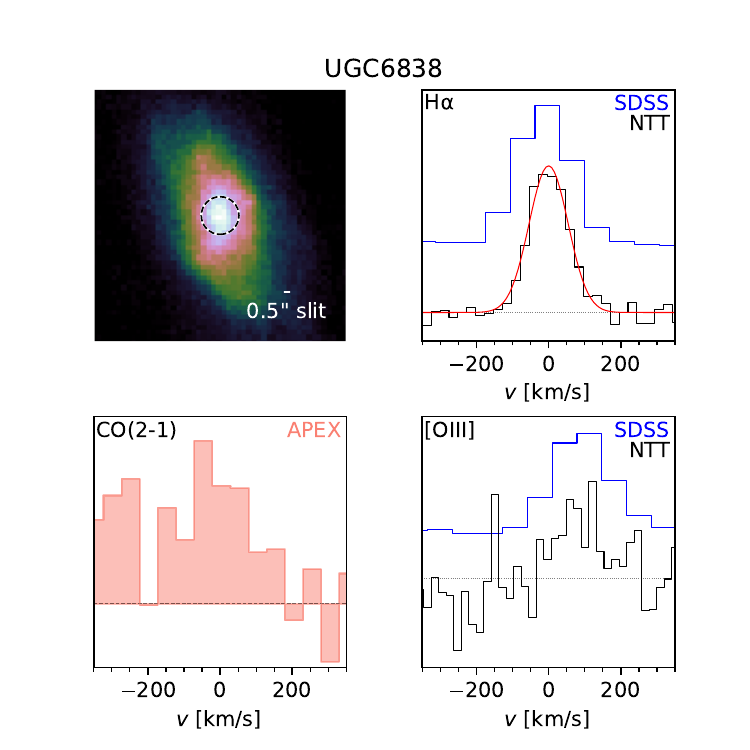}
\includegraphics[clip=true,trim=0.8cm 0 0.8cm 0.3,width=0.32\textwidth]{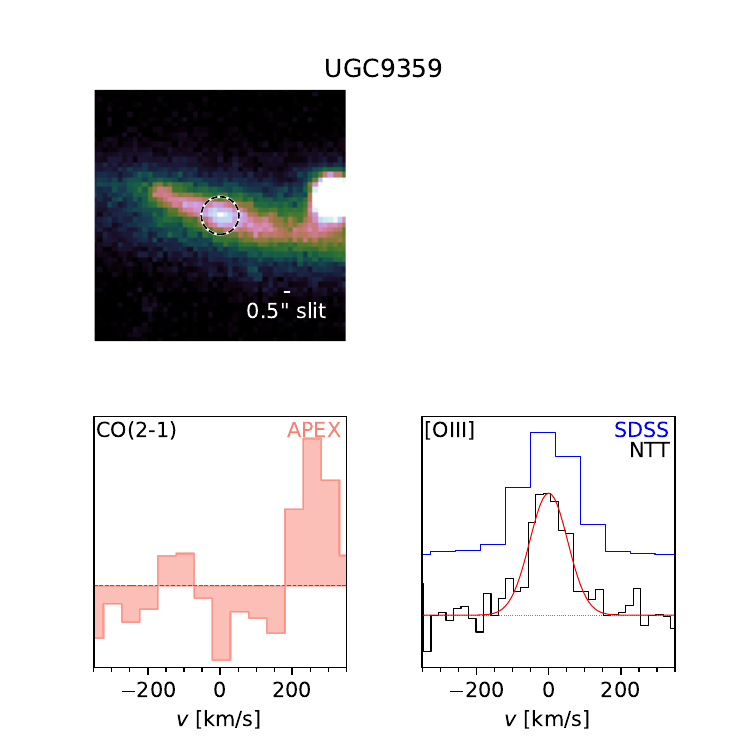}
\includegraphics[clip=true,trim=0.8cm 0 0.8cm 0.3,width=0.32\textwidth]{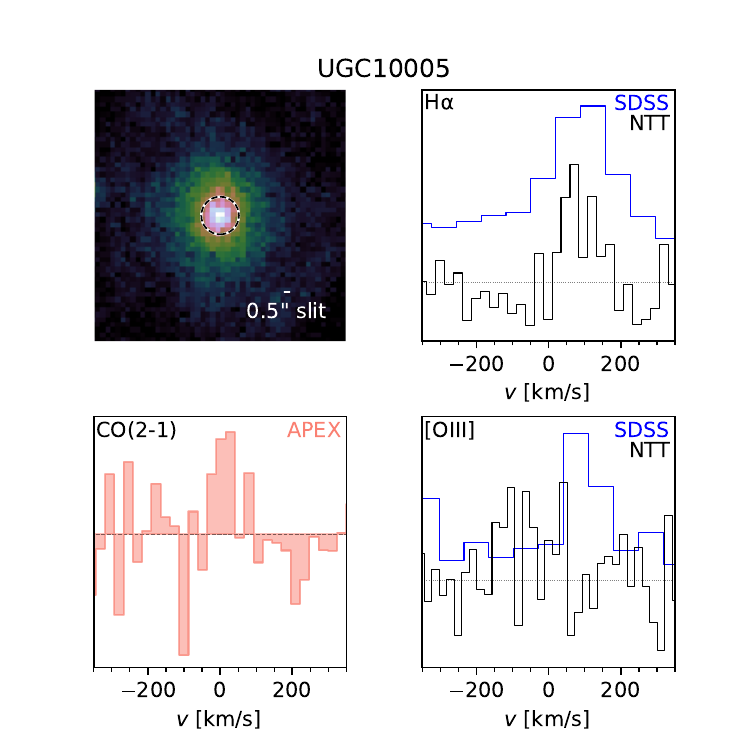}
\includegraphics[clip=true,trim=0.8cm 0 0.8cm 0.3,width=0.32\textwidth]{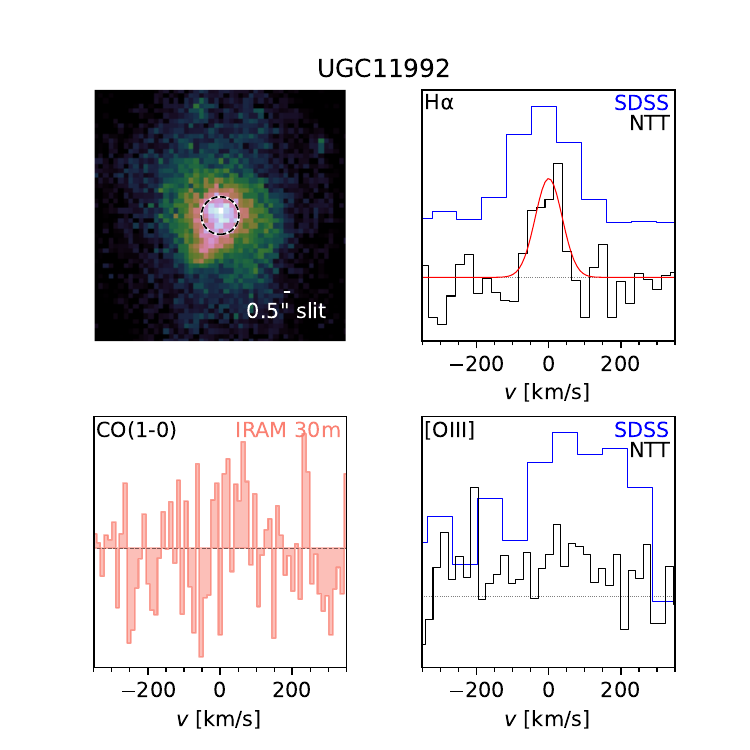}
\includegraphics[clip=true,trim=0.8cm 0 0.8cm 0.3,width=0.32\textwidth]{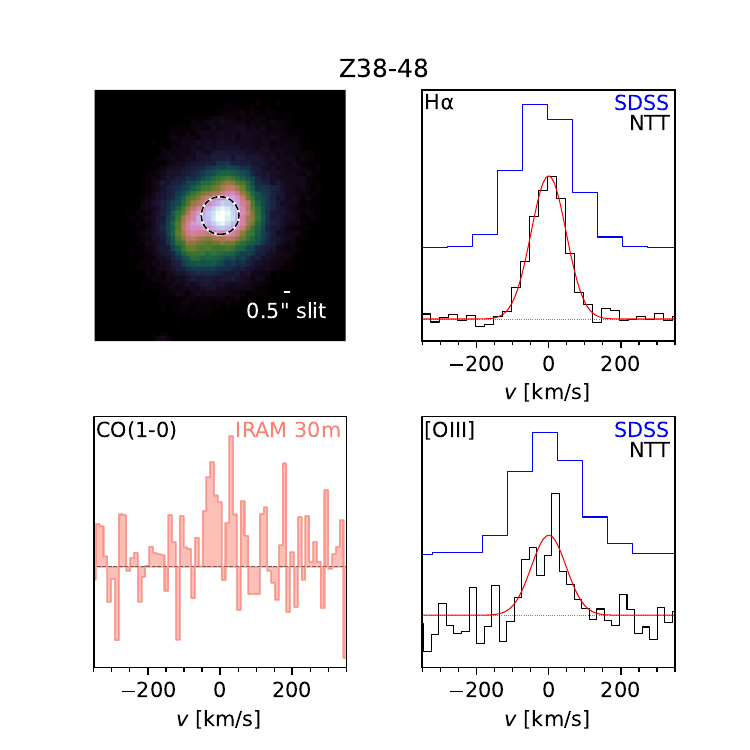}
\includegraphics[clip=true,trim=0.8cm 0 0.8cm 0.3,width=0.32\textwidth]{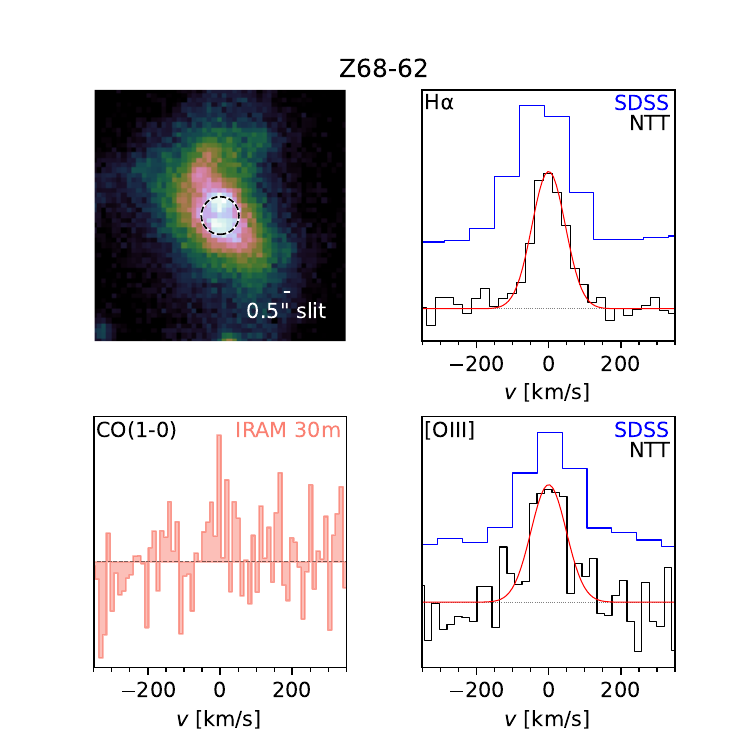}
\includegraphics[clip=true,trim=0.8cm 0 0.8cm 0.3,width=0.32\textwidth]{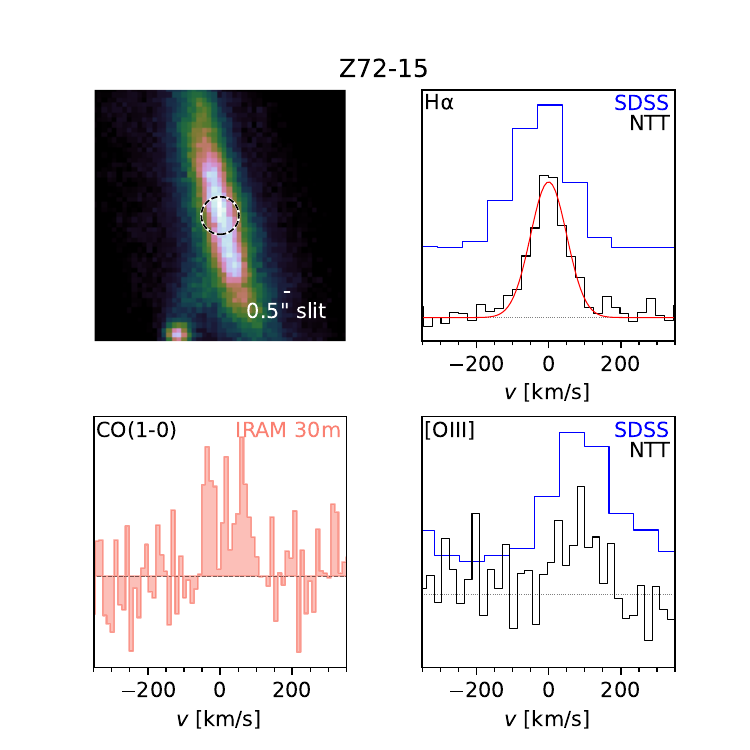}
\includegraphics[clip=true,trim=0.8cm 0 0.8cm 0.3,width=0.32\textwidth]{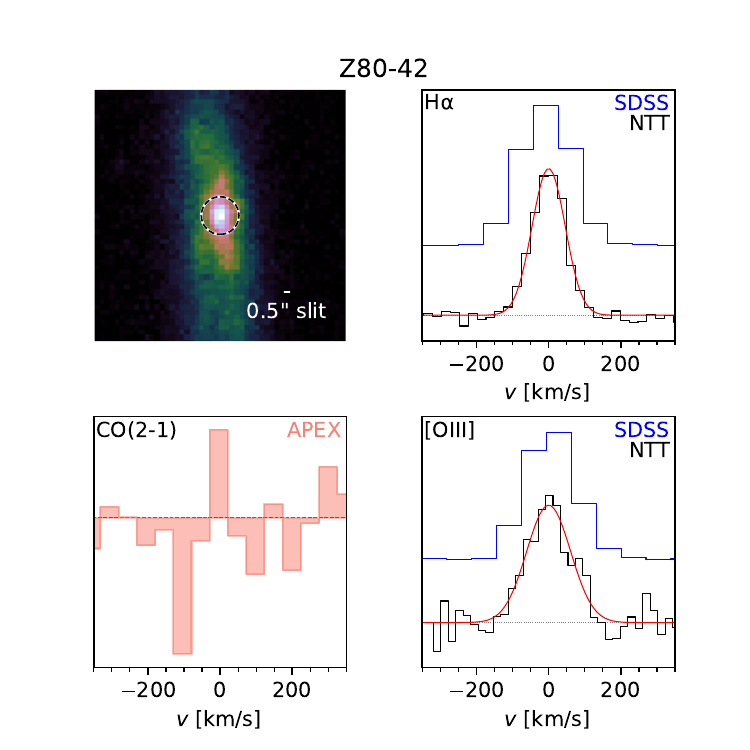}
\includegraphics[clip=true,trim=0.8cm 0 0.8cm 0.3,width=0.32\textwidth]{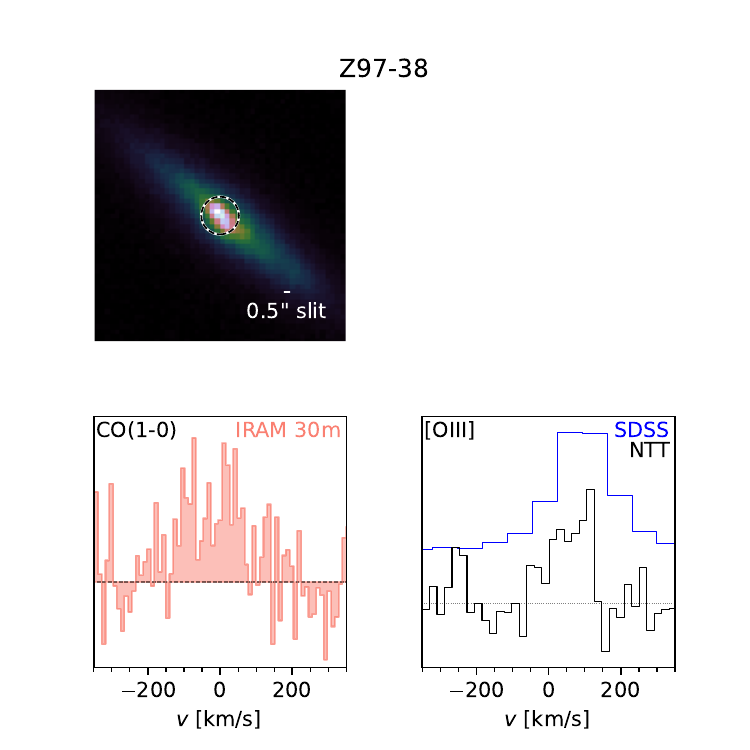}
    \caption{Fig.~\ref{fig:ha_cards_ntt_1} cont.}
    \label{fig:ha_cards_ntt_4}
\end{figure}

\onecolumn
\begin{figure}
    \centering
\includegraphics[clip=true,trim=0.8cm 0 0.8cm 0.3,width=0.32\textwidth]{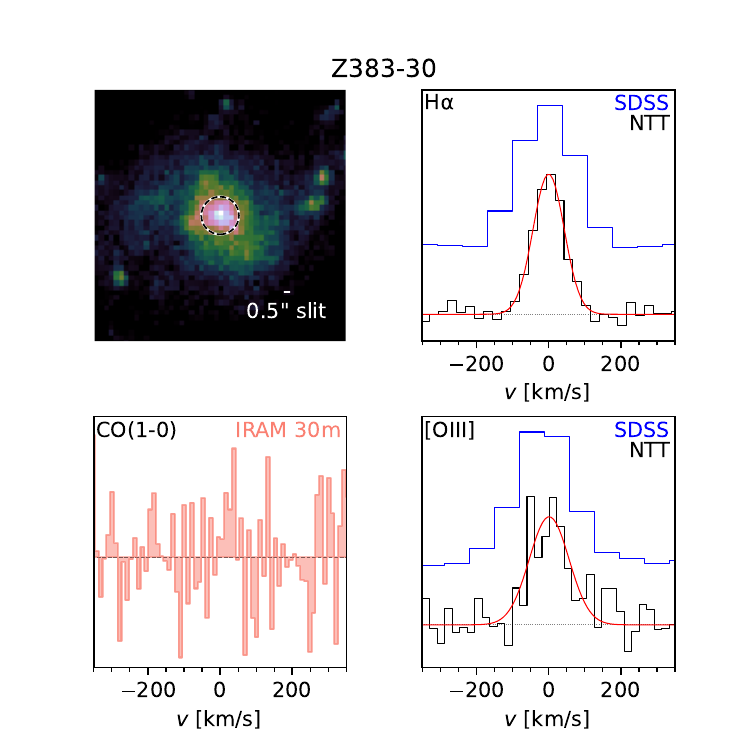}
\includegraphics[clip=true,trim=0.8cm 0 0.8cm 0.3,width=0.32\textwidth]{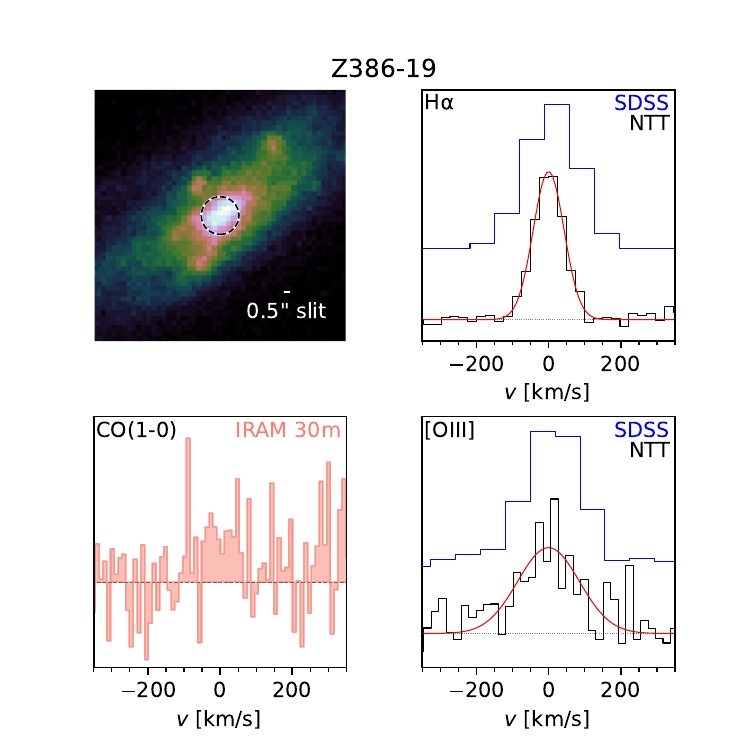}
\includegraphics[clip=true,trim=0.8cm 0 0.8cm 0.3,width=0.32\textwidth]{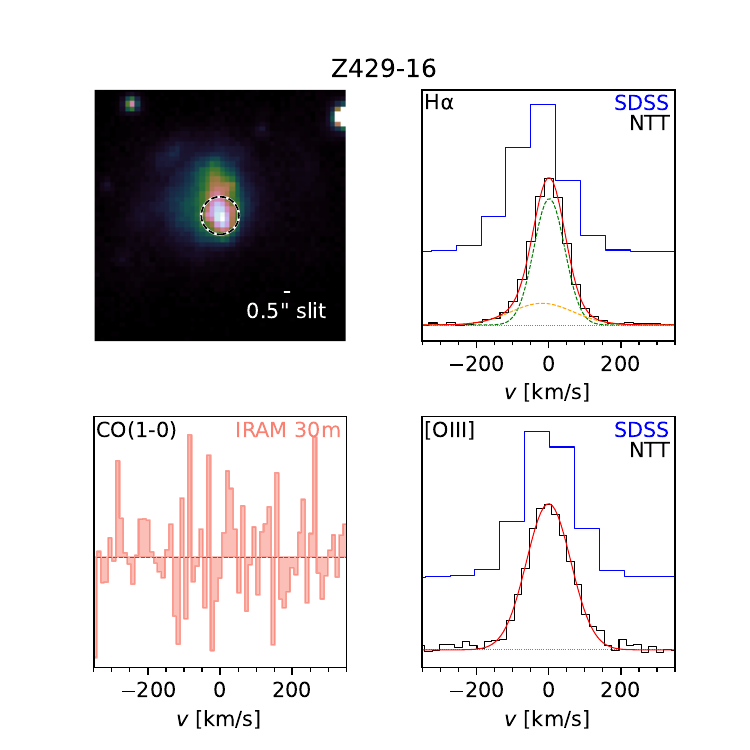}
\includegraphics[clip=true,trim=0.8cm 0 0.8cm 0.3,width=0.32\textwidth]{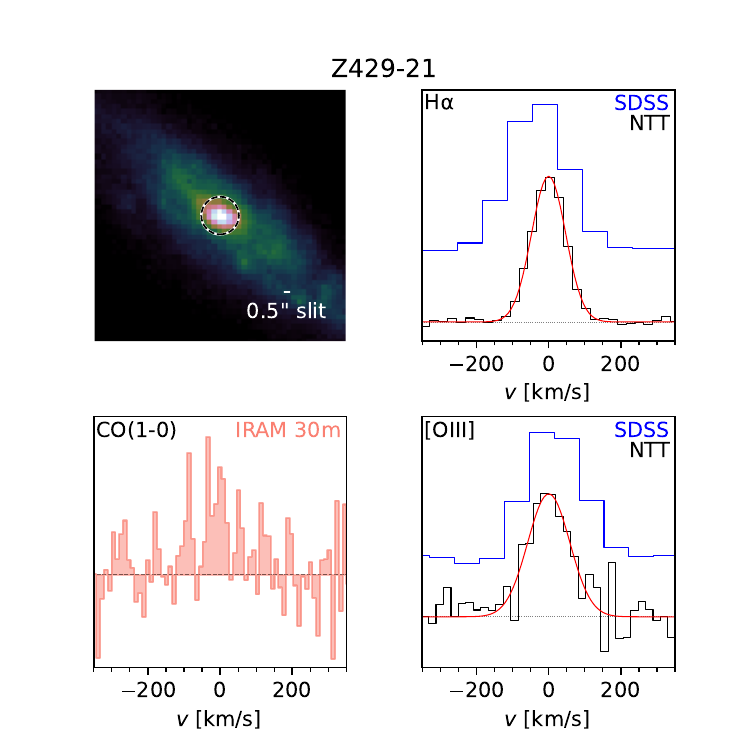}
\includegraphics[clip=true,trim=0.8cm 0 0.8cm 0.3,width=0.32\textwidth]{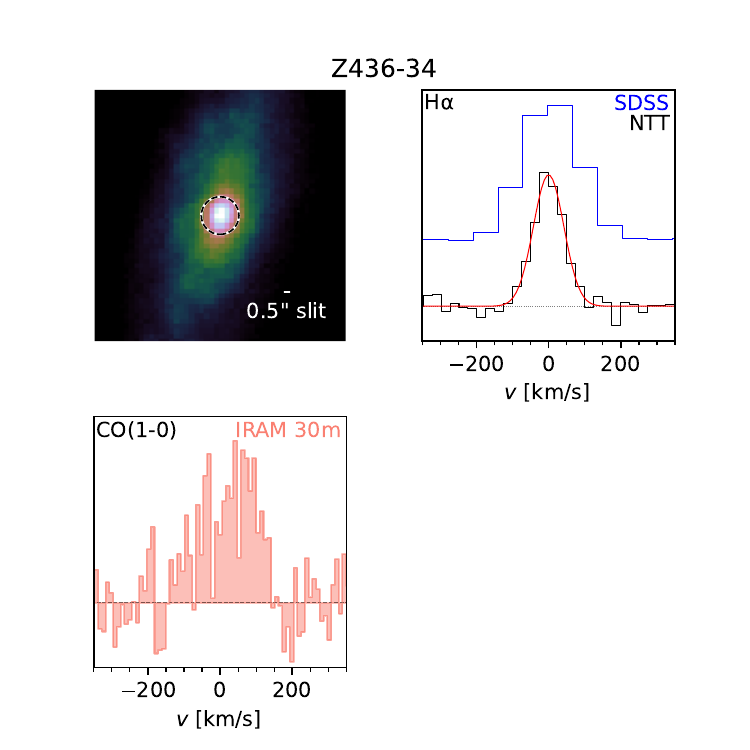}
    \caption{Fig.~\ref{fig:ha_cards_ntt_1} cont.}
    \label{fig:ha_cards_ntt_5}
\end{figure}

\onecolumn
\section{TNG/DOLORES line spectra}
\label{app:tng_specs}

\begin{figure}[h]
    \centering
\includegraphics[clip=true,trim=0.8cm 0 0.8cm 0.3,width=0.32\textwidth]{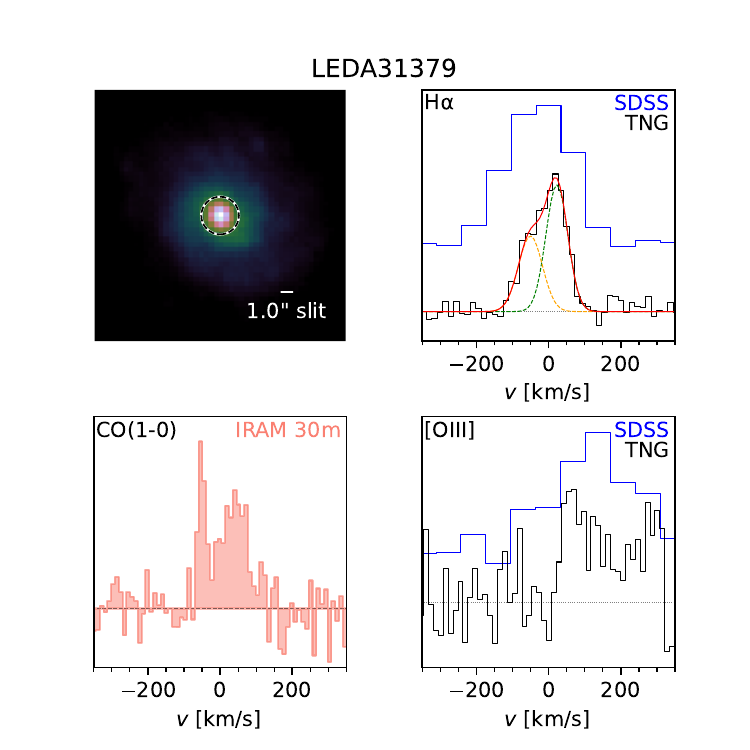}
\includegraphics[clip=true,trim=0.8cm 0 0.8cm 0.3,width=0.32\textwidth]{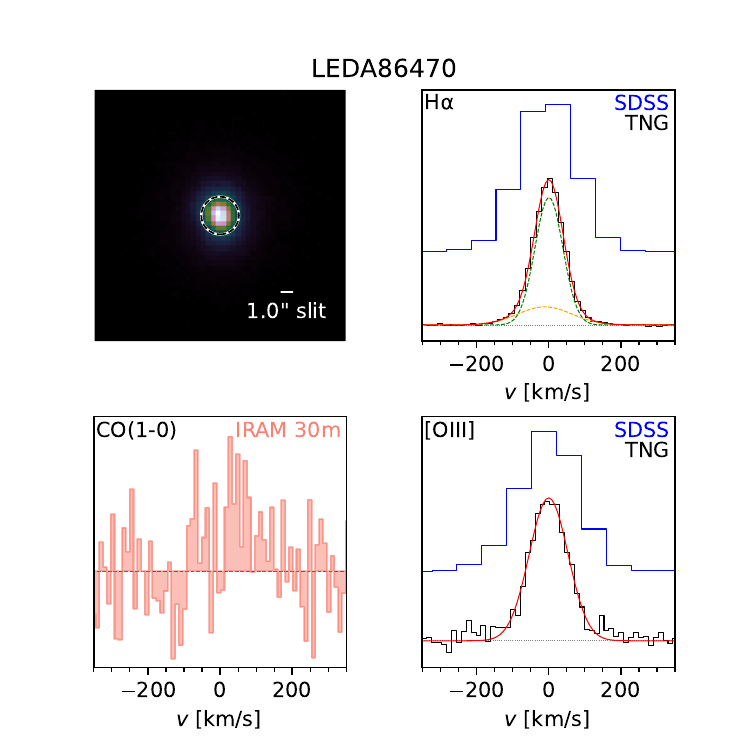}
\includegraphics[clip=true,trim=0.8cm 0 0.8cm 0.3,width=0.32\textwidth]{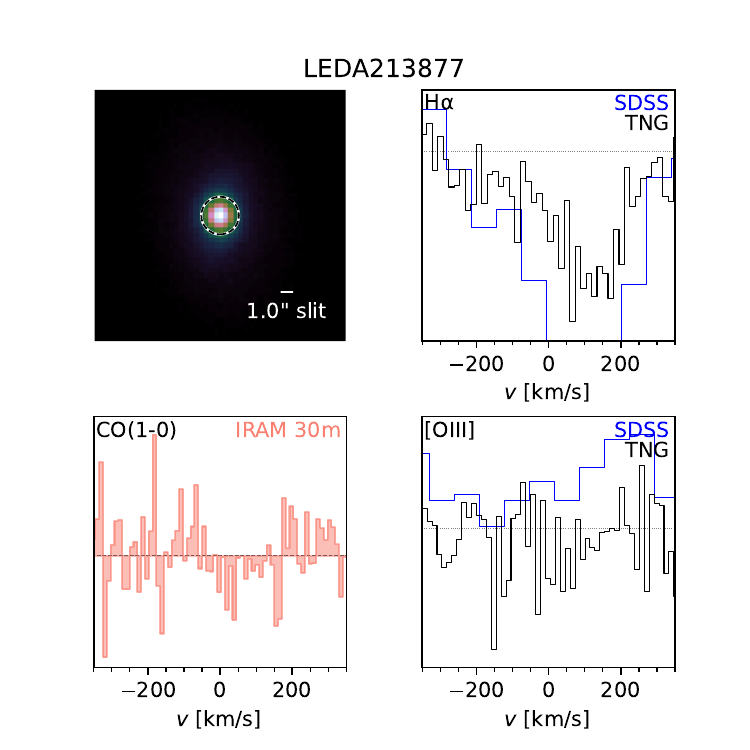}
\includegraphics[clip=true,trim=0.8cm 0 0.8cm 0.3,width=0.32\textwidth]{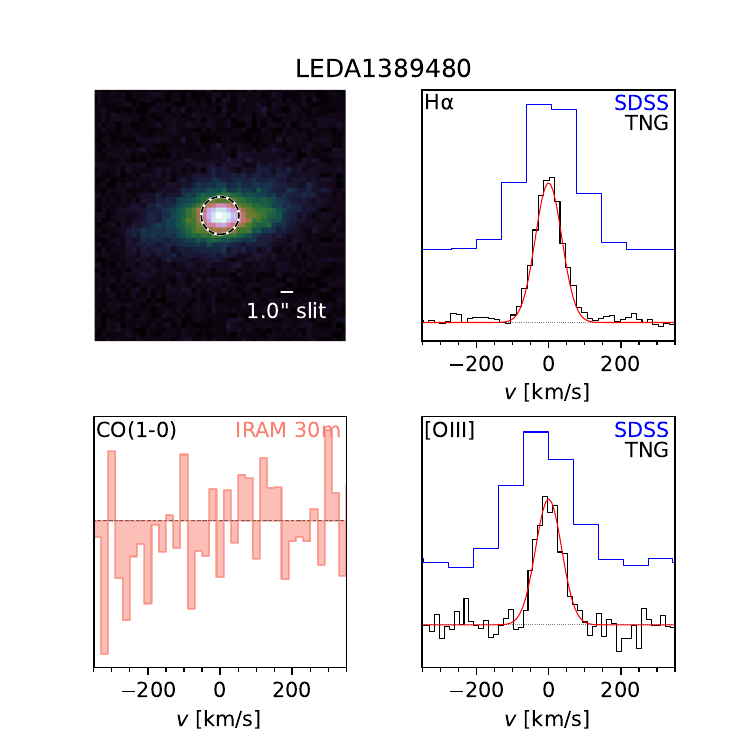}
\includegraphics[clip=true,trim=0.8cm 0 0.8cm 0.3,width=0.32\textwidth]{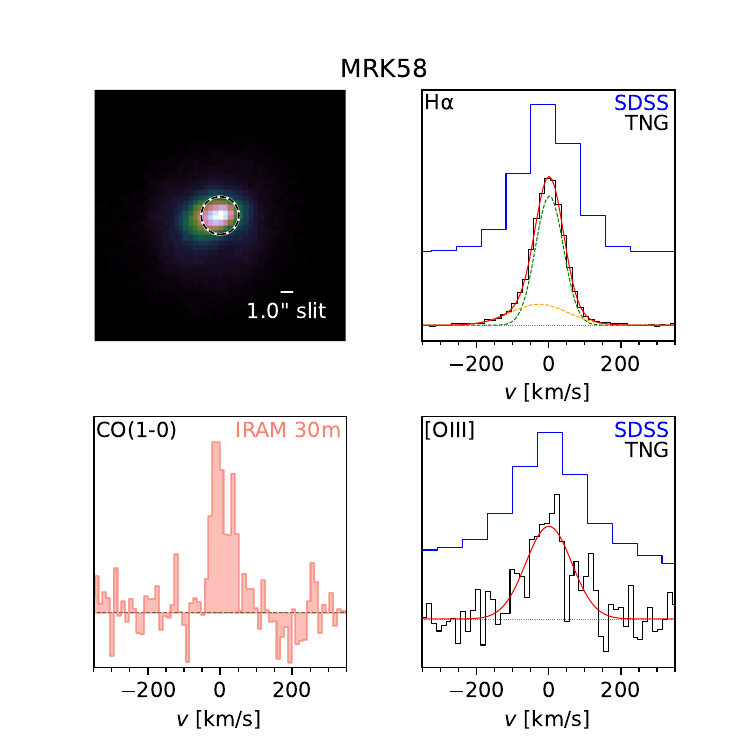}
\includegraphics[clip=true,trim=0.8cm 0 0.8cm 0.3,width=0.32\textwidth]{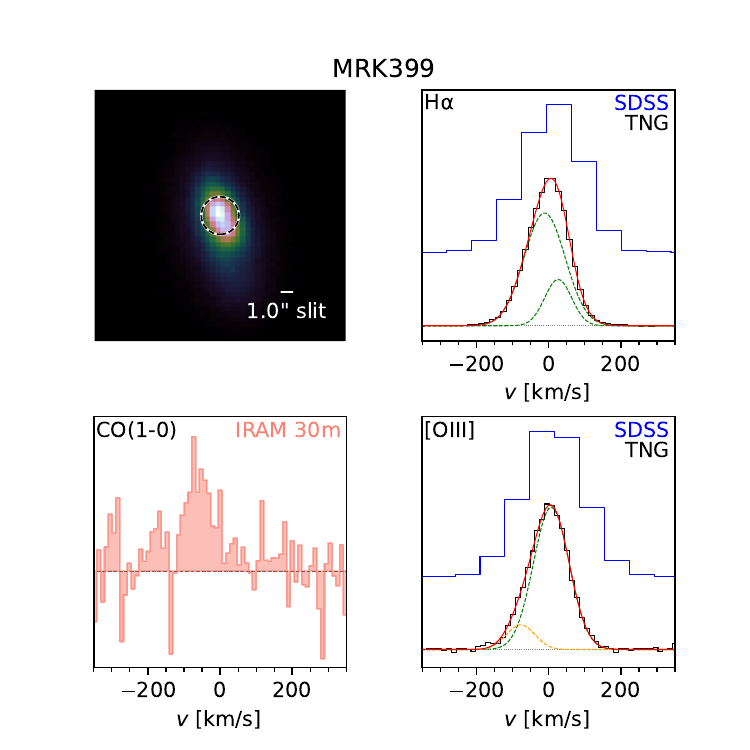}
    \caption{Targets observed with TNG/DOLORES. 
    The image on the top left in each panel shows the SDSS r-band image of the target, with the size and position of the 3'' SDSS fiber shown as a black-and-white dashed circle. 
    The white bar in each image shows the size of the slit used in the TNG/DOLORES spectroscopy.
    On the top right in each panel we show the SDSS (on top in blue) and higher resolution TNG/DOLORES (below in black) spectra around the H$\alpha$ emission line at 6562.8\,\AA, both converted to the same rest-frame velocity units. 
    On the bottom right in each panel is the same plot for the [OIII] emission line at 5006.77\,\AA.
    For emission lines detected with S/N$>$3 we also show the best fit model in red.
    If the model has multiple components, individual components are shown as dashed curves in green (non-outflow components) and orange (outflow components).
    On the bottom left is the region around the CO(1-0) or CO(2-1) emission line, whichever was used in the analysis, from the ALLSMOG and xCOLDGASS surveys.
    This figure is continued below in Fig.~\ref{fig:ha_cards_tng_2}.}
    \label{fig:ha_cards_tng_1}
\end{figure}

\onecolumn
\begin{figure}
    \centering
\includegraphics[clip=true,trim=0.8cm 0 0.8cm 0.3,width=0.32\textwidth]{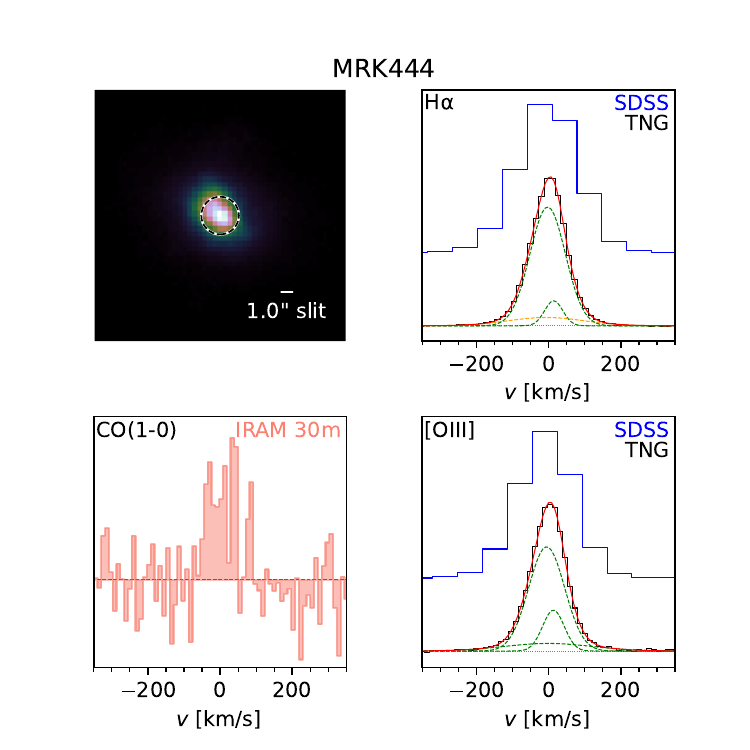}
\includegraphics[clip=true,trim=0.8cm 0 0.8cm 0.3,width=0.32\textwidth]{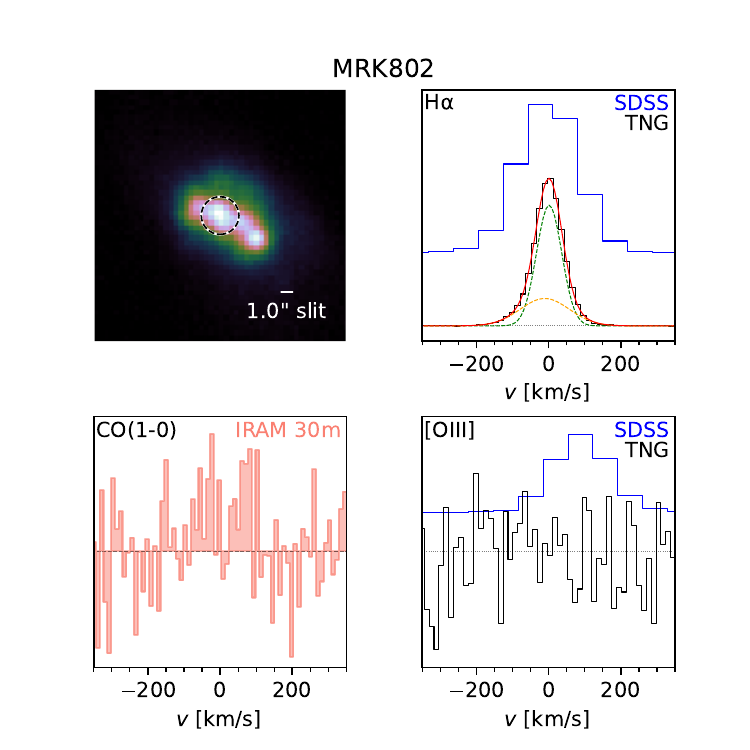}
\includegraphics[clip=true,trim=0.8cm 0 0.8cm 0.3,width=0.32\textwidth]{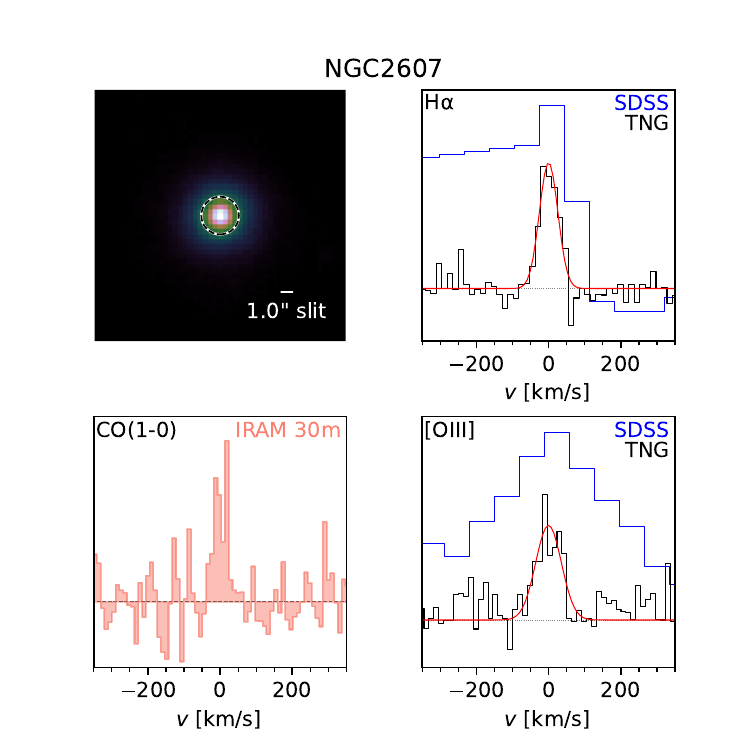}
\includegraphics[clip=true,trim=0.8cm 0 0.8cm 0.3,width=0.32\textwidth]{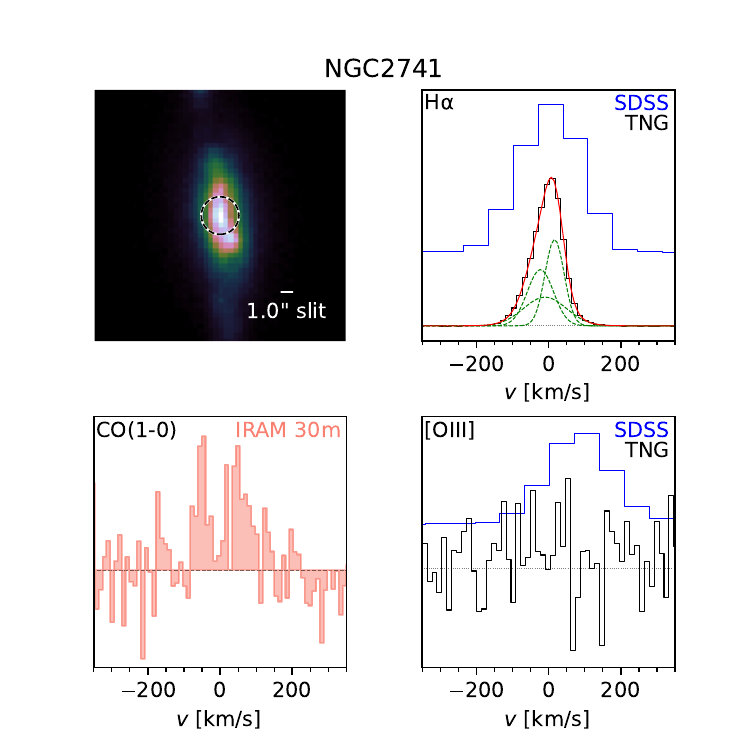}
\includegraphics[clip=true,trim=0.8cm 0 0.8cm 0.3,width=0.32\textwidth]{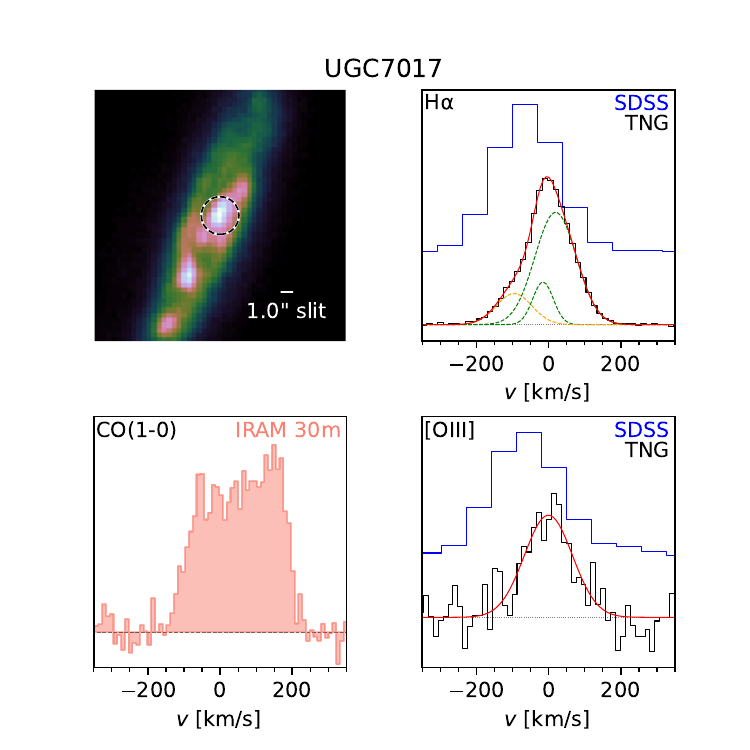}
\includegraphics[clip=true,trim=0.8cm 0 0.8cm 0.3,width=0.32\textwidth]{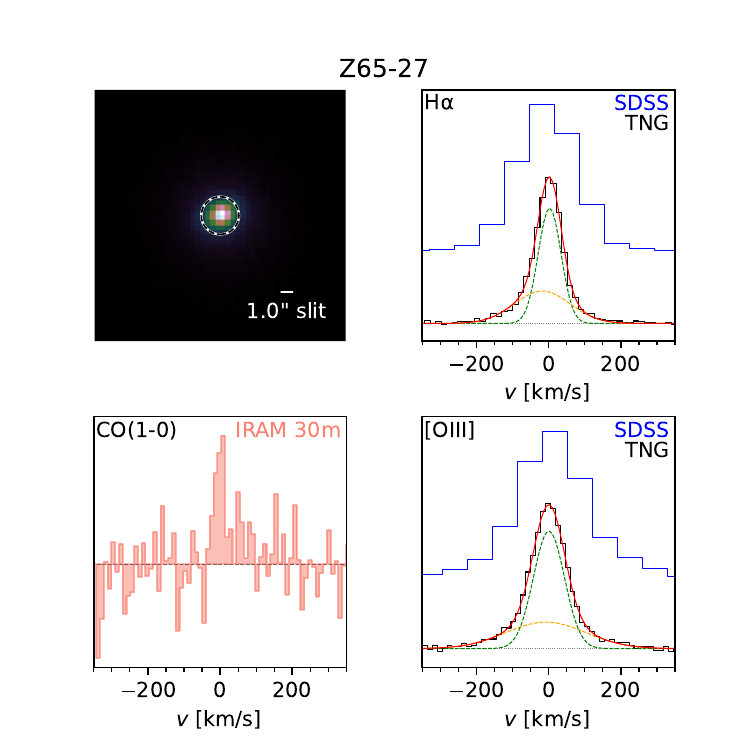}
\includegraphics[clip=true,trim=0.8cm 0 0.8cm 0.3,width=0.32\textwidth]{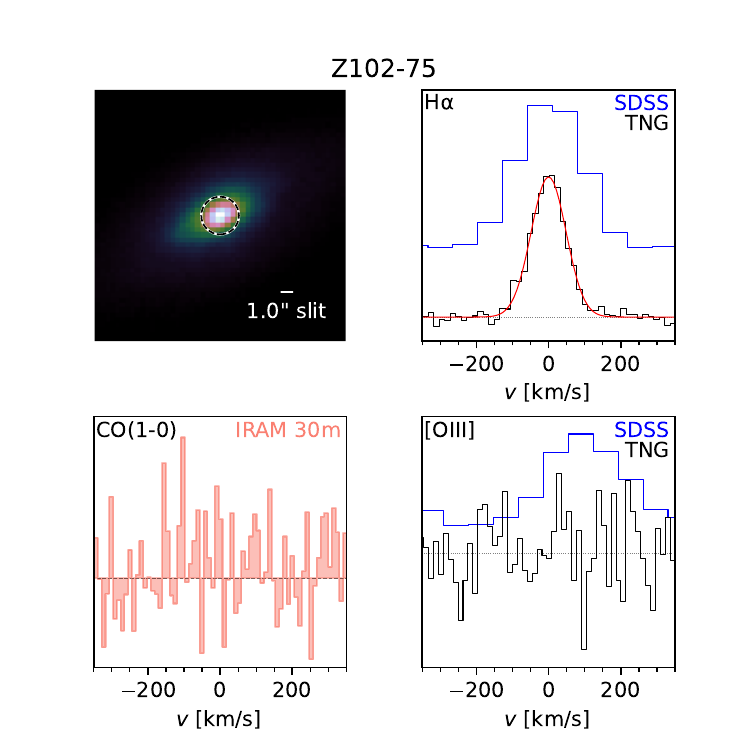}
    \caption{Fig.~\ref{fig:ha_cards_tng_1} cont.}
    \label{fig:ha_cards_tng_2}
\end{figure}

\end{document}